\RequirePackage{fix-cm}
\documentclass[twocolumn]{svjour3}          
\smartqed  
\usepackage{graphicx}
\usepackage{booktabs} 
\usepackage{rotating}
\usepackage{balance}
\usepackage{graphicx,graphics}
\usepackage{float}
\usepackage{multirow}
\usepackage{amsmath}
\usepackage{amssymb}
\usepackage[noend]{algpseudocode}
\usepackage{subfigure}
\usepackage[ruled,vlined]{algorithm2e}
\usepackage{latexsym}
\usepackage{colortbl}
\usepackage{booktabs}
\usepackage{comment}
\usepackage{array}
\usepackage{epstopdf}
\usepackage{balance}
\usepackage{adjustbox}
\usepackage{graphicx}
\usepackage{color}
\usepackage{amsmath}
\usepackage{booktabs}
\usepackage{amssymb}
\usepackage{color}
\usepackage{graphicx,graphics}
\usepackage{latexsym}
\usepackage{threeparttable}
\usepackage{colortbl}
\usepackage{makecell}
\usepackage{bigstrut}
\usepackage{mathrsfs}
\usepackage{lscape}
\DeclareMathAlphabet{\mathcal}{OMS}{cmsy}{b}{n}
\DeclareMathAlphabet{\mathcal}{OMS}{cmsy}{m}{n}

\definecolor{mygray}{gray}{.5}
\definecolor{mypink}{rgb}{.99,.91,.95}
\definecolor{mycyan}{cmyk}{.3,0,0,0}

\begin{document}

\title{Modern Data Pricing Models: Taxonomy and Comprehensive Survey}



\authorrunning{Xiaoye Miao et al.}

\author{Xiaoye Miao\textsuperscript{1}         \and
        Huanhuan Peng\textsuperscript{1}         \and
        Xinyu Huang\textsuperscript{1} \and
        Lu Chen\textsuperscript{2} \and \\ 
        Yunjun Gao\textsuperscript{2} \and
        Jianwei Yin\textsuperscript{1, 2}
}

\institute{Xiaoye Miao \at
              \email{miaoxy@zju.edu.cn}
        \and
          Huanhuan Peng \at
              \email{hhpeng@zju.edu.cn}
          \and
          Xinyu Huang \at
              \email{huangxinyu@zju.edu.cn}
         \and
          Lu Chen \at
              \email{luchen@zju.edu.cn}
        \and
          Yunjun Gao \at
              \email{gaoyj@zju.edu.cn}
        \and
          Jianwei Yin \at
              \email{zjuyjw@cs.zju.edu.cn}
          \at
\begin{description}
	\item[\textsuperscript{1}] Center for Data Science, Zhejiang University, Hangzhou, China
	\item[\textsuperscript{2}] College of Computer Science, Zhejiang University, Hangzhou, China
\end{description}
}

\date{Received: date / Accepted: date}

\maketitle

\begin{abstract}
Data play an increasingly important role in smart data analytics, which facilitate  many data-driven applications.
The goal of various data markets aims to alleviate the issue of ``isolated data islands'', so as to benefit \emph{data circulation}.
The problem of \emph{data pricing} is indispensable yet challenging in data trade.
In this paper, we conduct a comprehensive survey on the modern data pricing solutions.
We divide the data pricing solutions into \emph{three major strategies} and \emph{thirteen models}, including \emph{query pricing strategy}, \emph{feature-based data pricing strategy}, and \emph{pricing strategy in machine learning}. It is so far the first attempt to classify so many existing data pricing models.
Moreover, we not only elaborate the thirteen specific pricing models within each pricing strategy, but also make in-depth  analyses among these models.
We also conclude five research directions for the data pricing field, and put forward  some novel and interesting data pricing topics.  
This paper aims at gaining better insights, and directing the future research towards practical and sophisticated pricing mechanisms for better data trade and share.
\end{abstract}

\section{Introduction}
\label{intro}

The data from various data generation sources have significant economic and social value, which play an important role in big data era.
Data are transforming science, business, and governance by making decisions increasingly and by enabling data-driven applications.
Due to many reasons such as separated systems, company regulations, and  data privacy,  mining data value suffers from the ``isolated data islands'' where the data scatter at different places.
For example, most of data owners are very unwilling to share their data.
It thus drives the emergence of modern data markets to bridge the gap between data owners and data buyers under the dilemma of ``isolated data islands''.

\begin{figure}[t]
\centering
\includegraphics[width=0.45\textwidth]{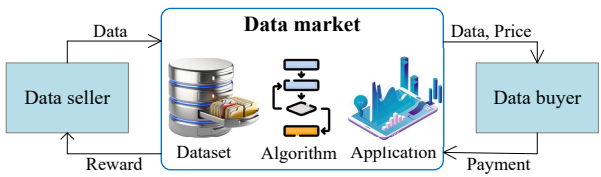}
\caption{The structure of a data market}
\label{fig:market}
\end{figure}

The data market acts as an intermediator for the buying and selling of diverse data, through which, the data generator (i.e., the seller) gets paid from the sold data, and the data consumer (i.e., the buyer) obtains beneficial data, as illustrated in Figure~\ref{fig:market}. In this way, the financial (resp. data) asset flows efficiently from the data consumer (resp. generator) to the data generator (resp. consumer).
A good data market is able to carry out the  data circulation among interested parties.
It not only allows groups to strategically purchase available datasets to avoid the labor costs and skill requirements necessary for data curation, but also helps them concentrate on how to
take advantage of these datasets and reliably generates
a profit from sought-after datasets.

As a consequence, data market has received much attention from both  \emph{industry} and \emph{academia}.
First, 
the emerging data markets (such as Microsoft Azure Marketplace~\cite{azure}, GBDEx~\cite{gbdex}, InfoChimps~\cite{infochimps}, etc.)  have the goals of completely enabling the data circulation and effectively solving the data isolation issue, so as to facilitate the openness and share of global data resources~\cite{youedata}.
On the other hand, regarding the academic research, the term ``data marketplace'' was firstly used by Armstrong and Durfee in 1998~\cite{armstrong1998mixing}, and then it attracts much attention~\cite{fernandez2020data,muschalle2012pricing,schomm2013marketplaces}.
Moreover, it is very common to trade data in real life for satisfying various needs, as revealed in the following examples.

\begin{example}\label{exm:estate}

\textbf{Pricing real estate trade data}.
Suppose a data company collects a real estate trade dataset $D$ in Hangzhou, a city of China.
This dataset not only contains the real estate price, but also records the information around the house, e.g., how many shopping malls around, the convenience of transportation facilities, etc.
It is beneficial for a \emph{real estate investment company} to make decisions of building a new apartment in certain region.
For example, an investment analyst, John, is willing to require a subset of the real estate trade dataset $D$, i.e., the real estate trade at West lake district of Hangzhou, to make a thorough quarterly report for this area.
Thus, in addition to trading the whole dataset, it is very common to trade the $D$'s subsets to meet various preferences.
On the other hand, for real estate investment, it is very important to obtain a real estate price prediction model for a newly developing district to accurately estimate the profit after investment.
In such situations, John concentrates more on some  machine learning models, instead of the original  dataset $D$.
Hence, there are urgent needs to trade the machine learning models.
\end{example}

\begin{example}\label{exm:flow}
\textbf{Pricing passenger flow data}.
Suppose there is a passenger flow dataset from commercial districts that is collected by the cameras in the malls.
The passenger flow data analysis is essential for managers to make smart business decisions and harvest more profit.
For instance, a restaurant manager, David, has to leverage the nearby passenger flow data on the second floor to  reasonably determine the marketing plan.
Hence, he needs to consume a fraction of the passenger flow dataset to achieve this goal.
Moreover,  predicting the number of passengers in the peak time of holidays  can better plan staff allocation in advance.
In this setting, it is necessary to trade machine learning models over the passenger flow dataset.
%
\end{example}
During data market operations, how to price the data is an indispensable problem.
Data pricing is the basis of data market, which determines the price of data trade.
An incentive pricing mechanism helps facilitate a dynamic and healthy data market development.
Actually, the problem of data pricing is important yet challenging in modern data markets, since the trading products are not limited to the raw data, which can be a part of data, machine learning models, etc.

Therefore, in this paper,  we conduct a comprehensive survey on data pricing solutions.
We build a taxonomy of data pricing solutions. It consists of \emph{three} pricing strategies,  i.e., the \emph{query pricing strategy} (QP strategy for short), and the \emph{feature-based data pricing strategy} (FP strategy for short), and \emph{pricing strategy in machine learning} (MLP strategy for short).
To the best of our knowledge, this is a thorough taxonomy of so many data pricing solutions, which  basically cover all the existing data pricing methods. 

Specifically, we elaborate the three pricing strategies and make in-depth analyses over them.
In particular, the query pricing strategy established by the database community aims to derive prices for diverse queries proposed by data buyers.
It is mainly composed of five pricing models, i.e., query pricing with data view (Q1), query pricing based on information value (Q2), query pricing using data provenance (Q3), auction-based query pricing (Q4), and pricing in mobile crowd sensing (Q5).
In contrast, the feature-based data pricing strategy focuses on how to price data with data features, such as data quality, data privacy, and so on.
There are four pricing models in this strategy, namely, data pricing based on quality (F1), personal data pricing using privacy (F2), personal data pricing with market value (F3), and pricing social network with influence (F4).
In addition, the pricing strategy in machine learning (ML) community  supports the trade of ML models.
To be more specific, there are four pricing mechanisms for pricing in machine learning, i.e., the noise-injection pricing mechanism (M1), pricing with shapley value (M2), end to end pricing framework (M3), and pricing in federated learning (M4).
Overall, these pricing schemes can be used to offer reasonable prices for various queries and ML models, so as to support the data pricing scenarios like Example \ref{exm:estate} and Example \ref{exm:flow}.   

Moreover, we identify five research challenges of data pricing, and provide some preliminary ideas on how to approach these challenges in principle.
We put forwards some interesting topics about data pricing in data markets for future direction.
We aim at providing useful insights for researchers who are interested in data pricing and data trade.
We do believe that, this survey puts an initial but substantial step to the development of data trade and share.
In addition, it is worthwhile to mention that, this survey is different from existing data pricing surveys, which either focus on  other products pricing such as broadband and digital products~\cite{liang2018survey,sen2013survey}, or only simply survey existing methods without offering a taxonomy~\cite{pei2020survey,Zhang2020survey}.
To the best of our knowledge, this is the first attempt to tackle the problem of taxonomy over data pricing.
In a nutshell, this paper mainly  has the following notable contributions.
%

\begin{itemize}
\item \textbf{New taxonomy.} We propose a new taxonomy of modern data pricing models. It consists of three parts, including query pricing (QP) strategy, feature-based data pricing (FP) strategy, and pricing strategy in machine learning (MLP).
\item \textbf{Comprehensive review.} We conduct a comprehensive overview of modern data pricing solutions to various scenarios. We elaborate representative data pricing models, and provide a series of insights and comparisons for modern data pricing models.
\item \textbf{Future directions.} We put forward five research challenges in terms of efficiency, universality, scalability, interdisciplines, and adaptability. We also present several future directions of data pricing study.
\end{itemize}

The rest of the paper is organized as follows. We introduce the basic concepts  of data markets in Section \ref{sec:concpet}.
Section~\ref{sec:taxonomy} establishes the taxonomy of data pricing solutions.
We elaborate three pricing strategies in Section \ref{sec:qpmodel}, Section \ref{sec:fbmodel}, and Section \ref{sec:mlmodel}, respectively.
Then, we highlight five challenges and some interesting topics over data pricing in Section~\ref{sec:future}.
Finally,  we conclude our survey in Section~\ref{sec:conclusion}. 

\section{Background}
\label{sec:concpet}

In this section, we describe the  data market structure. Then, we introduce some commercial data markets.

\subsection{Data Market Structure}






The  deployment of a typical data market is depicted in Figure~\ref{fig:market}.
Data market acts as a platform on which anybody (or at least a great number of potentially registered clients) can upload and maintain data sets \cite{fernandez2020data,ge2005model,muschalle2012pricing,north2010multiscale,schomm2013marketplaces}. Access to and use of the data are regulated through varying licensing models, and the trusted transactions are guaranteed.
The most common agents of data market are the data market itself, data sellers, and data buyers.
To be more specific, the data market acts as a broker between data buyers and data sellers, and takes necessary measures to make the deal smooth among them.
In particular, it utilizes different techniques to match the buyers' needs, and it is responsible for managing and pricing data. 
In the whole process, the data, algorithms, and applications are traded in the market.
The interaction between forces of demand-supply and the pricing signals they generate calls \emph{market dynamics}.
The data flow between sellers and buyers makes the data become assets.

The price of data is expected to not only reflect the valuation of data buyer, but also associate with the subsequent reward for data seller, and thereby to promote data share for a healthy data market.
As a result, the problem of data pricing is essential for data circulation.
However, it is  challenging due to the characteristics of data commodities.
Specifically, data pricing is actually a general concept. It could be with respect to the raw datasets,  further data analysis, or high-level data services.
Some typical data pricing cases are described below.
(i) Pricing datasets provided by many enterprises (e.g., Data Society \cite{datasociety}) is the basic pricing task for data markets.
(ii) How to set the price for a certain subset of data based on buyers' requirements becomes indispensable for various data analyses.
(iii) The data pricing also involves the high-level data services with advanced techniques or platforms, such as artificial intelligence, cloud computing, and so forth.

\begin{figure*}[t]
\centering
\includegraphics[width=0.95\textwidth]{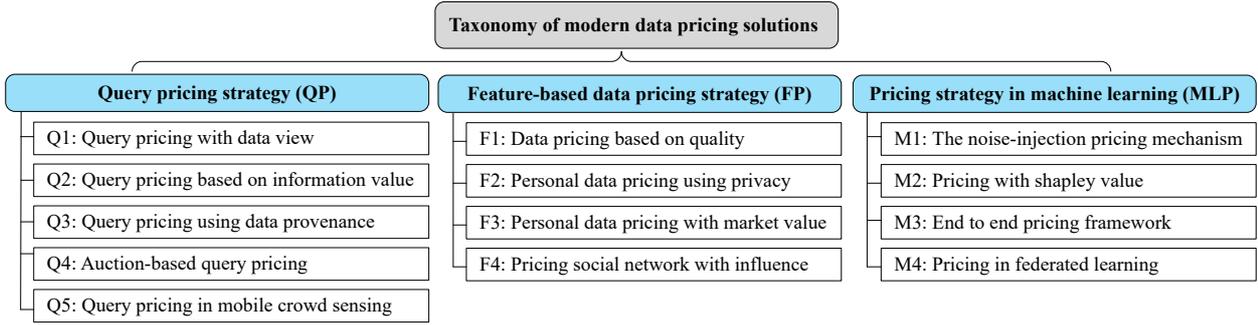}
\caption{Taxonomy of data pricing solutions}
\label{fig:taxonomy}
\end{figure*}

\subsection{Commercial Data Markets}
\label{subsec:cdm}
Existing commercial data markets sell  data either of multiple types or  in a specific field.

In particular, GBDEx~\cite{gbdex} possesses 225 high-quality data sources and more than 4,000 tradable data products, which has made 150PB trade volume involving more than thirty application fields. Microsoft Azure Marketplace~\cite{azure}, offers online data-driven applications and services either built on or designed to integrate with Microsoft's Azure public cloud.
InfoChimps~\cite{infochimps} is for sharing and selling data of all types with APIs access to 2,000 ``tables'' of data across several categories, including science, government, sports, and music.
Free datasets can be accessed up to 100,000 calls per month, and then, InfoChimps offers tiered pricing.

Meanwhile, there are a variety of data  markets selling data of specified types. For example, GNIP~\cite{gnip} aggregates and sells social media data from Twitter before the general data protection regulation (GDPR); Xignite~\cite{xignite} vends real-time financial data; Here~\cite{here} trades tracking and positioning data for location-based advertising; Factual~\cite{factual} enables location data to power innovation in product development, mobile marketing, and real-world analytics.
In addition, the value of human data is estimated at \$1.3 trillion but most of it is untapped. Datasift~\cite{datasift} helps extract and create more value in the ever-expanding universe of human data.

As summarized in data market surveys~\cite{liang2018survey,muschalle2012pricing,schomm2013marketplaces},
some data markets adopt \emph{freemium} pricing strategy, and the data can be used at no charge.
For example, Data.gov~\cite{datagov} is free as it is a website of the U.S. government.
Some data markets price data based on how many data are used in the service under the \emph{usage-based} pricing strategy.
In contrast, some data markets utilize the \emph{package} pricing strategy, that offers a customer a certain
amount of data or API calls for a fixed fee. Alternatively, after paying a fixed amount of money, customers can make unlimited use of the service for a limited time, mostly a month or a year (e.g., the typical subscription model).
Also, someone uses the \emph{flat fee tariff} pricing strategy  with minimal transaction costs, and it is based on time as the only parameter.

It is worthwhile to note that, the above pricing strategies are too simple to price complicated data.
First, the above strategies are based on several simple pricing methods for physical or digital goods, which are hard to support complex scenarios of data pricing.
For instance, they are not suitable to the scenarios described in Example~\ref{exm:estate} and Example~\ref{exm:flow}, due to the \emph{various} data pricing requirements.
Furthermore, they do not consider the core of data pricing, i.e., how it reflects the \emph{true value} of data.
Hence, the above pricing strategies still focus on the physical characteristics of goods, instead of the \emph{unique characteristics} of data.
Last but not least, there are many \emph{critical concerns} for data trade that are not considered, such as the privacy, the arbitrage, the revenue maximization, etc.
As a result, the further \emph{modern} data pricing models exclusively focus on the data pricing problem, which not only benefits diverse data trade, but also helps the data trade more reachable in real world.

\section{Taxonomy of Modern Data Pricing Models}
\label{sec:taxonomy}

In this section, we propose a new taxonomy of modern data pricing solutions, as depicted in Figure~\ref{fig:taxonomy}.

\begin{table*}[tbp]
\setlength{\tabcolsep}{3pt}
  \centering
  \caption{Comparison of Three Data Pricing Strategies}
\begin{tabular}{|c|c|c|c|c|}\hline
 \textbf{Name} & \textbf{Category} &  \textbf{Field}& \textbf{Pricing goal} & \textbf{Price subject} \\\hline
 QP & query pricing    &DB  & price diverse queries & the queries over data  \\\hline
FP & feature-based data pricing  & DB    & trade dataset based on data features &  the original data or processed data   \\\hline
MLP & pricing in machine learning   &  ML   & build a data market for ML &  the training data or ML model itself   \\\hline
\end{tabular}
  \label{tab:Mod}%
\end{table*}



This taxonomy consists of three pricing strategies and thirteen pricing models.
First, there are three strategies, i.e., the \emph{query} pricing (QP) strategy, the \emph{feature-based} pricing (FP) strategy, and the \emph{machine learning}  pricing (MLP) strategy.
These pricing strategies stand from different viewpoints.
In particular, the QP strategy allows data buyers to carry out different queries on datasets, in order to obtain desirable information.
The recent studies focus on how to derive price for diverse queries.
In contrast, the FP strategy prices data based on some features, such as data privacy, accuracy, etc.
While pricing in machine learning (w.r.t. the MLP strategy) aims to reasonably price data or models in machine learning.
The prices of models and data are derived according to machine learning related criterion, such as model accuracy.

Specifically, each pricing strategy contains a series of pricing models.
There are five, four, and four pricing models in the QP, FP, and MLP strategies, respectively.
The further classification on each strategy is based on the insights behind each pricing model, i.e., what the pricing model considers.
In other words, each model sets price from different perspectives, and thereby several necessary pricing issues are considered, such as arbitrage, market,  etc.

To begin with, in the QP strategy, the model of pricing with data view (i.e., Q1) aims to derive the query price when prices of several views (queries) are specified.
In particular, it requires the final price function satisfying a necessary property, i.e., \emph{arbitrage-free}, which means the data buyer cannot purchase a query by buying a group of other queries at a lower price.
This pricing model is feasible to deploy in real life, since it can support the pricing of diverse queries.
In contrast, the pricing model based on information value (i.e., Q2) trades a query over data according to the information value that the query brings.
Q2 has an assumption that, the data buyer has some prior knowledge of this database, i.e., a large set of possible database instances, which can be regarded as the candidate dataset containing the real dataset.
The information value measures how much information learned by the data buyers through the query, i.e., the uncertainty reduction  on candidate datasets.
This pricing model does not rely on the specified price points by human, which is hard to be influenced by immature human experiences, and thus it is practical to the case with less human knowledge.

Moreover, the pricing model using data provenance (i.e., Q3) assigns a query price  based on the sum price of used data (i.e., the lineage tuples) by the query.
The advantage of Q3 is the \emph{interpretability} of the query price, and meanwhile human can adjust the query price by changing the price of each datum.
Another equally important model is the auction-based query pricing model (i.e., Q4), which employs an auction to trade queries.
The queries in Q4 are treated as \emph{limited} goods, due to the limited  database resources and time constraints of data buyers.
Besides, this is the first auction mechanism designed for pricing a query (with limited resources).
This pricing model is more suitable for complex queries on big data, where the query cost cannot be ignored.
In addition, the fifth pricing model in QP (i.e., Q5) makes crowd sensing data trade applicable.
It models the mobile crowd sensing data as a distribution to support three spatial queries on the distribution.  
What's more, multi-versions of data goods are provided with the discounts based on the distances of versions.
The online pricing scheme is used to determine the final price with the purpose of maximizing the revenue.

In the second place, the FP strategy derives the data price according to some data features.
In particular, the quality-based pricing model (i.e., F1) considers the data quality as the pricing basis, such as accuracy, completeness, etc.
The lower the quality, the less the price.
This pricing model is simple, while it might be impractical to complicated pricing scenarios in real life.
In contrast, there are two kinds of personal data pricing methods  based on data features (i.e., F2 and F3).
The pricing model F2 trades the processed data, instead of the raw data, for privacy protection. 
It derives the price based on the privacy loss of all involved data.
Different from  F2, the pricing model F3 derives the optimal price that is close to the market value, in order to maximize the revenue.
Particularly, the market value is modeled as a function of data features, and an online learning fashion is employed to derive better price.
Besides, another pricing model in FP is pricing social network (w.r.t. F4).
It links the prices of nodes and their influence in social network together. 
This price setting is suitable for influence based applications, such as viral marketing and online advertising.

Last but not least, the MLP strategy consists of four pricing models.
The first one is the noise-injection pricing mechanism (i.e., M1), where the multi-version models with different accuracies are derived by adding different levels of noise.
Then, the prices of models are obtained to maximize the revenue.
In contrast, the pricing model M2 employs the shapley value to fairly distribute the revenue among data owners (sellers), i.e., pricing data among multi-parties.
In particular, the whole ML model is viewed as a cooperative game, and each seller is a player.
As a result, the marginal contribution of each datum is quantified by shapley value, and the price is assigned based on data's contribution.
Different from M1 and M2, the pricing model M3 takes both model and protected data into account.
It uses an end to end pricing framework to price data by the shapley value and privacy, i.e., the contribution and privacy of data.
Also, several versions of models are trained by a subset of data, satisfying the coverage rate and privacy parameter.
Based on market survey, the revenue maximization problem is proposed to obtain the optimal prices for different versions.
In addition, the pricing model in federated learning (i.e., M4), has to consider a series of factors (e.g., data size, transmission strategy, etc.) in order to incentivize the data share.

In summary, we compare the three pricing strategies from several aspects, as shown Table \ref{tab:Mod}.
In particular, the QP strategy aims to price queries over data, such as SPJ queries and linear query, for the database (DB) field.
Meanwhile, the FP strategy trades the original dataset or processed data according to the features of data, w.r.t. the database field.
For instance, the noisy aggregate statistics are priced due to the concern of privacy.
By contrast, the MLP strategy aims to price training data and models from the machine learning (ML) perspective, and thus to build a practical market for machine learning field.
In the following sections, we compare a series of representative pricing models in the three pricing strategies.
\section{Query Pricing}
\label{sec:qpmodel}

In this section, we first elaborate  five pricing models in the query pricing (QP) strategy, and then we compare them from several aspects.

Data buyers are often interested in extracting a proportion of  specific information from a dataset, as stated in  Example~\ref{exm:estate} and \ref{exm:flow}.
Access to this information can be concisely accomplished through a \emph{query}. However, the simple pricing schemes mentioned in Section~\ref{subsec:cdm} cannot directly support such scenarios. 
As a result, inspired by these, the query pricing (QP) strategy is introduced and studied in database community.
Since data buyers usually pursue the minimum monetary cost or  cannot afford the whole dataset,
they are willing to purchase the subsets of data that they need.

In the following, we are going to introduce the five representative pricing models in the QP strategy one by one, including Q1---the pricing model with \emph{data view}, Q2---the pricing model based on \emph{information value}, Q3---the pricing model with \emph{data provenance}, Q4---the \emph{auction} based pricing model,  and Q5---the pricing model for \emph{mobile crowd sensing}.

\vspace*{-0.1in}
\subsection{Query Pricing with Data View}
\label{subsec:view-model}
\vspace*{-0.041in}

The pricing model with data view (i.e., Q1) allows the seller to set the prices of a group of base queries (also called views), and it then derives the price of any new query  made to the database based on these views.

The Q1 pricing model is a simple pricing model and very \emph{practical} for query pricing problem.
It requires sellers to attach prices on specific queries, i.e., price points.
Once the prices of a set of popular views are fixed, any query is priced relying on these views' prices.
The views are like the \emph{anchor points} in the pricing model.
Existing studies of the Q1 pricing model  include pricing generalized chain queries~\cite{koutris2015query},
pricing SQL queries~\cite{koutris2012querymarket,koutris2013toward}, pricing aggregate queries~\cite{li2012pricing,wang2018pricing},   pricing general queries~\cite{deep2016design,lin2014arbitrage}, etc.

In this pricing model, the price function that assigns prices for diverse queries cannot be arbitrary.
In other words, the price function should be arbitrage-free.
Specifically, the \emph{arbitrage} property  is based on an \emph{instance based determinacy} relationship~\cite{koutris2012querymarket,koutris2013toward,koutris2015query}, as stated in Definition~\ref{defn:determinacy}.
Consider the USA business dataset: if $p$ is the price for the entire dataset and $p_1, \cdots, p_{50}$ are the prices for the data in each of the 50 states, a rational seller would ensure that $p_1 + \cdots + p_{50}\geq p$. Otherwise, no buyer would pay for the entire dataset, but would instead buy all 50 states separately. In general, a price function is said to be arbitrage-free if whenever a query $q$ is ``determined'' by the queries $q_1, \cdots, q_n$, their prices satisfy the inequality $p \leq p_1 +\cdots + p_n$.


Formally, let a query bundle $\mathbf{Q}$ = $(Q_1, \cdots, Q_n)$ be a finite set of queries that is asked simultaneously on the database. Given two query bundles $\mathbf{Q}_1$ and $\mathbf{Q}_2,$ we denote their union as $\mathbf{Q}$ = $\mathbf{Q}_1, \mathbf{Q}_2$. This $\mathbf{Q}$ is the query bundle consisting of all queries in $\mathbf{Q}_1$ and $\mathbf{Q}_2$.

\begin{definition}\label{defn:determinacy}
{\bf (Instance based determinacy)}~\cite{koutris2015query}.
Let $D$ be a database instance and $\mathbf{V}, \mathbf{Q}$ be two query bundles. It is said that, $\mathbf{V}$ determines $\mathbf{Q}$ given $D$, in notation $D\vdash\mathbf{V}\twoheadrightarrow \mathbf{Q}$, if for any $D'$, $\mathbf{V}(D) = \mathbf{V}(D')$ implies $\mathbf{Q}(D) = \mathbf{Q}(D')$.
\end{definition}

Informally, a query bundle $\mathbf{V}$ (that contains a set of queries/views)  determines a query $Q$ if one can compute answers of $Q$ only from answers of these views in $\mathbf{V}$ without having access to the underlying database. Specifically, given a database $D$, $\mathbf{V}$ determines $Q$
(denoted as $D\vdash\mathbf{V} \twoheadrightarrow Q$) if one can answer $Q$ from answers of $\mathbf{V}$ by applying a function $f$ such that $Q(D)$ = $f(\mathbf{V}(D))$~\cite{koutris2015query}. The impact on pricing is that if the user needs to answer the query $Q$,  he/she  has also the option of querying $\mathbf{V}$ and then applying $f$.
Put it differently, $\mathbf{V}$  determines $Q$ if $\mathbf{V}$ provides enough information to uniquely determine answers to $Q$.
If $\mathbf{V}$ determines $Q$, a potential buyer interested in purchasing $Q$ can
buy $\mathbf{V}$ instead, and derive the answers to $Q$ from $\mathbf{V}$: arbitrage \emph{occurs} when the price of $\mathbf{V}$ is lower than that of $Q$.

\begin{definition}\label{defn:arbitrage-freedom}
({\bf Arbitrage-freedom})~\cite{koutris2015query}.
 A price function $p_D$ is arbitrage-free if, whenever $D\vdash\mathbf{Q}_1, \cdots, \mathbf{Q}_k\twoheadrightarrow \mathbf{Q}$, then $p_D(\mathbf{Q})\leq \sum_{i=1}^kp_D(\mathbf{Q}_i)$.
\end{definition}
\begin{proposition}
\label{pro:free-qpricing}
Any arbitrage-free price function $p_D$ satisfies the following properties. In particular, $p_D()$ denotes the price of the empty bundle, and the identity bundle, denoted by $\mathbf{ID}$, is the bundle that returns the entire dataset $D$.

\noindent
$(i)$ Subadditivity: $p_D(\mathbf{Q}_1, \mathbf{Q}_2)\leq p_D(\mathbf{Q}_1) + p_D(\mathbf{Q}_2)$.

\noindent$(ii)$ Nonnegativity: $p_D(\mathbf{Q})\geq 0$.

\noindent$(iii)$ Not asking is free: $p_D()=0$.

\noindent$(iv)$ Upper-boundedness: $p_D(\mathbf{Q})\leq p_D(\mathbf{ID})$.
\end{proposition}


The first property exactly refers to \emph{bundle arbitrage}. It regards the scenario where a data buyer that wants to obtain the answer for the bundle $\mathbf{Q}$ = $\mathbf{Q}_1$, $\mathbf{Q}_2$ creates two separate accounts, and uses one to ask for $\mathbf{Q}_1$ and the other to ask for $\mathbf{Q}_2$.
To avoid such an arbitrage situation, we must make sure that the price of $\mathbf{Q}$ is at most the sum of the prices for $\mathbf{Q}_1$ and $\mathbf{Q}_2$.

Following these definitions, the QueryMarket system automatically prices a large class of SQL queries requested by buyers based on price-points specified by sellers~\cite{koutris2012querymarket,koutris2013toward}.
Since computing arbitrage-free prices is theoretically hard, a novel approach that translates them into optimized integer linear programs (ILPs) is proposed for computing prices.
The idea is to find the minimum sum price of views that cover the query results over database.
It also supports updates to the database and accounts for query history. In addition, it allows multiple sellers to share revenues fairly.

On the other hand, there is a similar study on the pricing of aggregate queries~\cite{li2012pricing}.
It assigns the prices to the special type of linear queries, namely the aggregate queries that can be expressed as a linear combination of the column entries of a relationship.
In particular, the seller prices a base set of linear queries, and then the market algorithm computes the price of the remaining queries such that it satisfies \emph{arbitrage freeness}.
Different from~\cite{koutris2015query}, the arbitrage freeness is based not on an \emph{instance based determinacy} relationship but rather on a \emph{schema based determinacy}: if for any instance of the database, one could determine a linear query $Q$ using the answer to a bundle of linear queries $Q_1, \cdots, Q_k$, then it must be that $p(Q) \leq \sum_{i=1}^k p(Q_i)$.
This crucial difference makes the price function independent of the database instance.
In addition, there is a theoretical framework~\cite{wang2018pricing} to support pricing \emph{approximate} aggregate queries in a more realistic scenario, where consumers are willing to accept the approximate aggregate query (due to the constraints on time, resources, etc.).
In particular,  a transforming function is used to convert the original arbitrage-free price function to the one that supports approximate aggregate queries.



Moreover, with the main goal of avoiding arbitrage, a formal framework for pricing queries over data allows the construction of general families of price functions~\cite{deep2016design,lin2014arbitrage}.
They consider two pricing schemes: \emph{instance-independent} scheme, where the price depends only on the structure of the query, and \emph{answer-dependent} scheme, where the price also depends on the query output.
Besides, it is worth exploring the relationship between arbitrage-freeness in the query pricing framework and envy-freeness in pricing theory for appropriately chosen buyer valuations~\cite{syrgkanis2015pricing}.


\subsection{Query Pricing based on Information Value}
\label{subsec:qp-infovalue}

The information value based pricing model Q2  derives the query price based on the information value of each query, i.e., how much information that the query brings.

It is important to note that, although Q2 allows sellers to set prices for specific queries, it is feasible when there is no price point setting.
Moreover, the specified price points may have negative impact on the price functions in Q1~\cite{lin2014arbitrage}.
Thus, there is no requirement on the pricing preparation and the seller knowledge in Q2, which is more suitable when data sellers or data markets have a little pricing knowledge.
Meanwhile, Q2 reflects the relative value of data, and thereby it is hard for human to change the derived prices.  Hence, it enjoys rationality when human experiences do not work.


The framework \textsc{Qirana}, standing the viewpoint of the data buyer, employs the \emph{possible databases} semantic to price relational queries~\cite{deep2017qirana,deep2017qirana-d}.
First, it is assumed that, the schema of the trading database $D$ is known to the data buyer, and he also has some knowledge about the dataset $D$.
That is to say, the data buyer has a set of possible database instances $\mathscr{D}$ (that includes $D$).
With the new bought query, one can easily find that, the possible database $\mathscr{D}$ that $Q(\mathscr{D})\neq Q(D)$ is not the real database $D$.
In other words, he can discard the possible database $\mathscr{D}$ that $Q(\mathscr{D}) \neq Q(D)$ through the bought query, and thus the candidate possible databases shrink.
In this way, the data buyer learns more \emph{information} about the real database $D$.
Hence, \textsc{Qirana} assigns a price to $Q$, which can be formulated as a function of how the possible databases shrinks,  i.e., how much information that the query brings.
Since it is infeasible to keep track of all the possible databases, it attempts to circumvent this issue by leveraging the \emph{support set} $\mathcal{S}$ that is a subset of all possible databases.
The support set is generated via a random neighborhood sampling method.

Given the support set $\mathcal{S}$ and the output of a query bundle $\mathbf{Q}(D)$ (i.e., a set of queries as mentioned in Section~\ref{subsec:view-model}), a data buyer knows a \emph{conflict set}, i.e., $\mathcal{C}_{\mathbf{S}}(\mathbf{Q},D)=\{\mathscr{D} \in \mathcal{S} \mid \mathbf{Q}(\mathscr{D}) \neq \mathbf{Q}(D)\}$.
Any database that $\mathscr{D} \in \mathcal{C}_{\mathbf{S}}(\mathbf{Q},D)$ is not the possible database any more, thus these databases can be removed from the database candidate set.
There are several price functions in \textsc{Qirana}, including \emph{weighted coverage pricing}, \emph{uniform entropy gain pricing}, etc.
In particular, the weighted coverage price function $p^{w c}(\mathbf{Q}, D)$ computes the price as the sum of assigned weights of disagreements, as defined in Eq. \ref{eq:wc}. 
While the uniform entropy gain price function $p^{u e g}(\mathbf{Q}, D)$ models the price as the gain in entropy,
as written in Eq.~\ref{eq:ueg}.
\begin{equation}\label{eq:wc}
p^{w c}(\mathbf{Q}, D)=\sum_{D_i \in \mathcal{C}_{\mathbf{S}}(\mathbf{Q},D)}w_{D_i} 
\end{equation}
\begin{equation}\label{eq:ueg}
p^{u e g}(\mathbf{Q}, D)=\frac{\log \left|\mathcal{C}_{\mathbf{S}}(\mathbf{Q},D)\right|}{\log |\mathcal{S}|}
\end{equation}

Furthermore,  \textsc{Qirana} supports the case that some price points $\left(\mathbf{Q}_{j}, p_{j}\right)$ are specified, by solving a entropy maximization problem.
It also optimizes the pricing computations, in order to support history-aware pricing, batch updates, and aggregation.
Finally, a pricing system for efficiently implementing \textsc{Qirana} is developed.
\textsc{Qirana} utilizes the weighted coverage price function by default, and it acts as a \emph{broker} between the buyer and the seller. More details on the price functions can refer to~\cite{deep2017qirana,deep2017qirana-d}.

Following this pricing framework, the revenue maximization problem on query pricing is also identified~\cite{chawla19revenue}.
The price functions are based on the information value, leveraging the concept of \emph{support set} $\mathcal{S}$.
Specifically, it conducts a hypergraph  $\mathcal{H}=(\mathcal{V}, \mathcal{E}),$ with vertex set $\mathcal{V}=\mathcal{S},$ and hyperedges $\mathcal{E}=\left\{e_{i} \mid i=1, \ldots, m\right\},$ where $e_{i}=\mathcal{C}_{S}\left(\mathbf{Q}_{i}, D\right)$, and $\mathcal{C}_{S}$ is the conflict set of $\mathbf{Q}_{i}$ with respect to $\mathcal{S}$.
Each vertex represents the possible database in the conflict set, while each query $\mathbf{Q}_{i}$ is denoted as a hyperedge $e_i$, containing the databases in its conflict set.
Accordingly, the revenue maximization task is casted as \emph{bundle pricing} problem for \emph{single-minded} buyers and \emph{unlimited} supply.

To be more specific, several simple price functions are proposed, including  uniform bundle pricing, item pricing, and $XOS$ pricing, with efficient price computation algorithms.
In particular, uniform bundle pricing assigns a same price $p_0$ to every hyperedge, i.e., $p^{b}(e) = p_0$.  While item pricing first assigns a weight $w_j$ for each vertex and then computes price as the weight sum of vertices in the hyperedge, i.e., $p^{a}(e)=\sum_{j \in e} w_{j}$. $XOS$ pricing first defines $k$ weights $w_j^{1},...,w_j^{k}$ for each vertex and uses the maximum weight sum of vertices in the hyperedge, i.e., $p^{x}(e)= \max _{i=1}^{k} \sum_{j \in e}  w_{j}^{i}$ \cite{chawla19revenue}.
When the price function is certain, a series of algorithms are employed to tackle the revenue maximization problem, and the final price is derived.

\vspace*{0.1in}
\subsection{Query Pricing using Data Provenance}
\label{subsec:provenance-model}

Given the prices of tuples in relations, the pricing model with the data provenance (i.e., Q3) prices the query based on the set of the tuples contributing to the result tuples of a query~\cite{miao2020towards,ruiming2014quality,tang2013price}.
Different from Q1 and Q2, the Q3 pricing model assigns the query price according to the data usage of the query (like the factors of production in microeconomics). This is a common way to price relational data in commercial data markets \cite{azure}.

Given a database with each tuple having a price, the price of a query in the Q3 pricing model is a function of the prices of tuples that contribute to the response of the query.
The data contribution/usage is measured via the concept of data lineage from the perspective of data provenance in the database field.
\emph{Provenance} information describes the origin and the history of data in its life cycle.
Each tuple $t$  occurs  in the output of a query with a set of tuples presented in the input, called the lineage of $t$~\cite{cui2000tracing}.
The lineage of $t$ is meant to collect all of the input data that ``contribute to'' $t$ or help to ``produce'' $t$, as stated in Definition~\ref{defn:min-lineage-tuple}.

\begin{definition}\label{defn:min-lineage-tuple}
{\bf (Tuple's lineage set)}.
Given a dataset $D$ with tables $T_1, \cdots, T_m$,  and a query  $Q$.
Let $Q(D)$ = $Q(T_1, \cdots, T_m)$ be the result set of the query $Q$ over tables $T_1, \cdots, T_m$.
For a tuple $t\in Q(D)$, one $t$'s lineage set w.r.t. $Q$  in $T_1, \cdots, T_m$, denoted as $L(t\in Q(D), D)$ $(L(t, D)$ for short$)$, is defined by Eq.~\ref{eq:minimal-tuple-lineage}.
\begin{equation}\label{eq:minimal-tuple-lineage}
L(t, D) = \bigcup_{i=1}^m T^*_i
\end{equation}
\begin{equation}\label{eq:tuple-lineage-vector}
Q^{-1}_{\langle T_1, \cdots, T_m\rangle}(t) = \langle T^*_1, \cdots, T^*_m\rangle
\end{equation}

\noindent
Eq. \ref{eq:tuple-lineage-vector} is the vector form of a lineage set of $t$, with each element $T^*_i$ having tuples from the table $T_i$.
For $i = 1, \cdots, m$, $Q^{-1}_{T_i}(t) =  T^*_i$ is $t$'s lineage in $T_i$, and each tuple in $T^*_i$ does contribute to the result tuple $t$.
Formally, $T^*_1, \cdots, T^*_m$ are subsets of $T_1, \cdots, T_m$ satisfying

\noindent(a) $Q(T^*_1,$ $\cdots,$ $T^*_m)$ $= \{t\}$;

\noindent(b) $\forall T^*_i, \forall T^{'}\subseteq T^*_i, Q(T^*_1,$ $\cdots,$ $T^{'}, \cdots, T^*_m)=\varnothing$.
\end{definition}

In fact, the condition (a)  constrains  that, the lineage tuple sets (i.e., $T_i^*$'s) derive exactly $t$, and the condition (b) indicates that, each tuple in the lineage indeed contributes something to $t$~\cite{cui2000practical,cui2000tracing}.

%
%
%
%
%
%
%

As a consequence, based on the query lineage, the Q3 pricing model is developed to accumulate the prices of the query lineage tuples \cite{miao2020towards,ruiming2014quality,tang2013price}.
The pricing solutions in \cite{ruiming2014quality,tang2013price}  fulfill desirable properties such as contribution monotonicity, bounded-price and contribution arbitrage-freedom, and support  query pricing for probabilistic databases.
In particular, the price of each probabilistic tuple is redefined with probability and price, and the query price is obtained.

In addition, the query pricing mechanism over incomplete data \textsf{iDBPricer}~\cite{miao2020towards}
consists of two designated price functions, i.e., the usage, and
completeness-aware price function (i.e., the UCA price) and the quality, usage, and completeness-aware price function (i.e., the QUCA price).
This pricing model prices the query based on the tuples that contribute to the query.
In particular, in addition to the data contribution,  the completeness and query quality are considered in \textsf{iDBPricer}, making the prices more practical.
It also supports to derive the history-aware prices.
It considers more market factors, and thus it provides opportunities for data sellers or markets to adjust prices based on market conditions.

\subsection{Auction-based Query Pricing}
\label{subsec:db-auction}

The query price is influenced by the value of data and  buyers' willingness, while the value of data is hard to define.
However, in a truthful auction, users are natural to submit their true valuations as the bids, which reflects the value of query to the bidder.
The pricing model based on auction mechanism  (i.e., Q4) is the first attempt to leverage  auction into query pricing~\cite{wang2019novel}.


In the auction based pricing model, the queries are modeled as \emph{limited goods}, which is different from the query pricing models Q1-Q3.
The former models Q1-Q3 have an implicit assumption that, the queries over data can be seen as unlimited supply.
In contrast, complex queries in large dataset are often costly to return approximate answers \cite{kandula2016quickr}.
Inspired by this, this pricing model Q4 employs the auction mechanism to price queries under constraints over the query cost (i.e., the resource demand and execution time).

In particular, the bidder in the auction submits a bid and the corresponding deadline for a query. He/she  only wins the bid when the resource demand of this query is within the database's capacity, and the estimated execution time is within the deadline.
To estimate the query cost, each query is decomposed, and the query plan is generated, in order to obtain the execution time and resource demand of the query.
For example, the resource demand such as the CPU cost in each time and the whole execution time are estimated.
Moreover, the auction is constructed, where a group  of coming bidders can submit as many bids as they like, but they could only win at most one bid~\cite{li2013designing,zhang2014dynamic}.

In this setting, a social welfare maximization problem is formulated as an integer linear programming (ILP)  problem, which considers the utility of bidders and the market (i.e., service provider) together.
This maximization problem is an optimization problem with the resource cost and execution time constraints.
To solve this NP-hard problem, it is first transformed to a dual linear programming problem.
Moreover, a greedy primal dual algorithm is proposed in both offline and online scenarios.
In the offline setting, the algorithm iteratively selects the ``best'' bid from the remaining bids, i.e., the maximum value of the radio between price and resource requirements.
For the online case, it directly checks any arriving bid whether the resource requirements are satisfied and accepts the bid with higher price.
Finally, a truthful and efficient auction mechanism is derived, which achieves near-optimal social welfare with a good approximate radio.

\begin{table*}[tbp]
  \setlength{\tabcolsep}{3pt}
  \centering
  \caption{Comparison of Query Pricing Models}
    \begin{tabular}{|c|c|c|c|c|c|c|c|}
    \hline
    \textbf{Model} & \textbf{Reference} & \textbf{Perspective} & \textbf{Mechanism} & \textbf{Presetting} & \textbf{Technique}  & \textbf{Arbitrage-free} & \textbf{Market} \\
    \hline
     Q1 & \cite{deep2016design,koutris2012querymarket,koutris2013toward,koutris2015query,li2012pricing,lin2014arbitrage,syrgkanis2015pricing,wang2018pricing} &  price inference & posted pricing & view price & ILP  & $\checkmark$    & $\checkmark$ \\
    \hline
     Q2 &\cite{chawla19revenue,deep2017qirana,deep2017qirana-d}     & information value & posted pricing & optional & sampling  & $\checkmark$    & $\checkmark$ \\
    \hline
     Q3 &\cite{miao2020towards,ruiming2014quality,tang2013price}    & data provenance & posted pricing & tuple price  & query lineage  & $\checkmark$    & $\checkmark$ \\
    \hline
     Q4 &\cite{wang2019novel}     &  limited supply & auction & N/A  & ILP  & uncertain   & $\checkmark$ \\
    \hline
     Q5 &\cite{zheng2017online,zheng2019arete}     &  data distribution & online pricing & basic price  & versioning  & $\checkmark$   & $\checkmark$  \\
    \hline

    \end{tabular}%
    \label{tab:pricing-qpmodels}
\vspace*{0.08in}
\end{table*}%

\subsection{Query Pricing in Mobile Crowd Sensing}
\label{subsec:db-crowd}

Leveraging the crowd power to acquire large-scale sensing tasks is called \emph{mobile crowd sensing} (MCS).

Armed with electronic items like smartphones, MCS has been widely explored and applied in many real-life scenarios.
With a smartphone, individuals are capable of collecting all kinds of  data around them. For example, global positioning system (GPS) in mobile phones can be utilized to collect traffic information and help users estimate the travel time, while the phone sensors can help track the individual behaviors to evaluate the impact on the environment pollution.
Therefore, MCS data are easy to collect and trade in real life.
%
%


The pricing problem in mobile crowd sensing data markets (w.r.t. Q5) is firstly identified in~\cite{zheng2017online}.
It is faced up with three challenges in MCS, including data uncertainty, economic-robustness (arbitrage-freeness in particular), and revenue maximization.
First, the powerful \emph{Gaussian process} regression technique is adopted to model the data uncertainty.
Thus,  the basic data commodity is formed as the posterior gaussian model that represents the data distribution in certain region $\Theta$.
In particular, for each location $y \in \Theta$, there is a random variable $\boldsymbol{X}_{y}$ associated with it.
As a result, a set of random variable $\boldsymbol{X}_{Y}$ is modeled for a set of locations $Y\subseteq\Theta$, and the probability density function for the joint distribution of $\boldsymbol{X}_{Y}$ is formed as follows, where $\boldsymbol{\mu}_{Y}$ and $\Sigma_{Y Y}$  denote the mean vector and the covariance matrix of $\boldsymbol{X}_{Y}$, respectively.
$$f\left(\mathbf{x}_{Y}\right)=\frac{1}{(2 \pi)^{|Y| / 2}\left|\Sigma_{Y Y}\right|^{1 / 2}} e^{-\frac{1}{2}\left(\mathbf{x}_{Y}-\boldsymbol{\mu}_{Y}\right)^{T} \Sigma_{Y Y}^{-1}\left(\mathbf{x}_{Y}-\boldsymbol{\mu}_{Y}\right)}$$

Since the possible locations in a certain region is infinite, the data consumer selects a set of locations in several locations, which is also called as point of interets (PoIs, denoted as $\mathbb{Y}$).
Moreover, three query types are provided to support query pricing in mobile crowd sensing, i.e., single-data query, multi-data query, and range query, so that the data consumers can obtain required information.
These queries are different from traditional SQL queries, since the data product is a posterior distribution.
The single data query asks for the mean value of a certain location $y_{i}$, the multi-data query answers the mean value vector in a certain region  $Y_{i}$, and the range query returns the probability that the data at a region $Y_i$ falls in a given range $\left[\underline{a}_{i}, \bar{a}_{i}\right]$.
In particular, the specific queries supported in the solution are formulated as follows.

\begin{itemize}
  \item \textbf{Single-data query}: A data consumer is interested in the (inferential) data at a single location $y_{i} \in \mathbb{Y},$ i.e., the (posterior) mean $\bar{\mu}_{y_i}$ of $y_{i}$.
  \item \textbf{Multi-data query}: A data consumer wants to know the (inferential) data of a certain region $Y_{i} \subseteq \mathbb{Y}$, i.e., the (posterior) mean vector $\overline{\boldsymbol{\mu}}_{Y_{i}}$ of $Y_{i} .$
  \item \textbf{Range query}: A data consumer asks for the probability that the data at the region $Y_{i} \subseteq \mathbb{Y}$ belongs to a range $\left[\underline{a}_{i}, \bar{a}_{i}\right]$.
\end{itemize}

 Moreover, the proposed pricing scheme \textsc{Arete} consists of two parts, i.e., versioning mechanism and an online pricing mechanism.
In the versioning phase, \textsc{Arete} generates different \emph{versions} of the basic conditional Gaussian distribution, where the distance of versions is measured through modified relative entropy (i.e., Kullback-Leibler distance).
The distance $\widehat{D}$ of two distributions $f_1$ and $f_2$ is formalized in Eq.~\ref{eq:kl}, where $\sigma_{1}$ and $\sigma_{2}$ are the variances of $f_1$ and $f_2$, respectively.
\begin{equation}\label{eq:kl}
\widehat{D}\left(f_{1} \| f_{2}\right)=\frac{1}{2}\left(\log \frac{\sigma_{2}^{2}}{\sigma_{1}^{2}}+\frac{\sigma_{1}^{2}}{\sigma_{2}^{2}}-1\right)
\end{equation}

To achieve a price with more profit, an online pricing mechanism is designed.
In particular, \textsc{Arete} determines the price from the price candidate set, in order to maximize the revenue.
In this setting, the price of basic version is required, since prices of other versions are proportional to basic price and the distance between two versions.
It is theoretically proved that, \textsc{Arete} is able to achieve both arbitrage-freeness and a
constant competitive ratio in terms of revenue maximization.

Moreover, based on \textsc{Arete}, \textsc{Arete-sh} is proposed to further share the reward, which can incentivize data providers to contribute data~\cite{zheng2019arete}.
In particular, the problem is formulated as a coalitional game with $m$ data providers and a reward vector.
First of all, \textsc{Arete-sh} defines the qualified basic \emph{coalition} by using versioning algorithm, in order to reduce the number of coalitions.
Since the basic \emph{coalition} is substitutable,  a compact representation scheme is employed, i.e., marginal contribution networks, to capture this feature.
In particular, marginal contribution networks can represent coalitional game by a set of rules with specific forms.
After that, shapley value is used to fairly share the reward in different coalitions.
As a result, \textsc{Arete-sh} can compute shapley value in polynomial time, and it achieves the four fairness axioms, including efficiency, symmetry, dummy, and additivity.

\subsection{Analysis on Query Pricing Models}
\label{subsec:db-analysis}

In this part, we compare these query pricing models Q1-Q5 from several aspects, as summarized in Table~\ref{tab:pricing-qpmodels}.


\textbf{Pricing perspective.}
The five query pricing models assign the query price from different angles.
The pricing model Q1 with data view  infers the query price from prices of several views.
As explicitly described,  Q2 assigns the query price via measuring the information value brought by the query, while Q3 uses data provenance to price the query according to the data used in query.
The pricing model Q4 employs an auction mechanism to obtain the query price, considering limited supply.
While the pricing model Q5 in mobile crowd sensing treats data distribution as the goods, and the price is derived based on the distance of different versions on data distribution.


\vspace*{0.042in}
\textbf{Pricing mechanism.} Most of the pricing models use the posted pricing strategy, e.g., Q1, Q2, and Q3.
In particular, the posted pricing scheme means that, the sellers/buyers post the prices they are willing to sell/pay, which is widely used in many application domains, such as selling flight tickets~\cite{gal2011pricing} and  products in Amazon~\cite{AmazonProduct},  cloud service trading~\cite{wang2012cloud}, and labor markets (e.g., crowdsourcing and crowdsensing)~\cite{hu2017optimal,qu2018posted}.
While the pricing model Q4 employs the auction mechanism to maximize social welfare.
For Q5 in mobile crowd sensing, it derives the optimal price through an online way.  Hence, it belongs to online pricing strategy.

\vspace*{0.042in}
\textbf{Presetting.}
Some pricing models require to set the base price or some constants in advance. It means that, some initial settings must be determined before pricing, which is a  preparation step of query pricing.
In QP, a query is priced depending on the predetermined prices.
Within the QP strategy, Q1 requires the price points (i.e., view price) specified by data sellers.
It is optional for Q2 to set the predetermined prices, since it is feasible when there is no predetermined price point.
In contrast, Q3 requires the price setting of tuples, in order to derive the final query price.
While the predetermined price is not applicable (i.e., not required) for Q4.
Besides, Q5 needs a price for the basic data version, based on which it determines the optimal price for revenue maximization in terms of different versions.

\vspace*{0.045in}
\textbf{Techniques of price computation.}
The efficiency of price computation is an important factor to evaluate a data pricing model.
Among the query pricing strategy, both Q1 and Q4 employ the integer linear programming (ILP) to solve the pricing problem.
The sampling technique is used in Q2 to reduce the number of possible databases, and thus improves  efficiency.
While Q3 utilizes the concept of query lineage to track data usage on the query, so as to derive the final query price.
The versioning technique is employed in Q5 to generate different versions of data for trade.


\textbf{Arbitrage-free property.}
The arbitrage freedom is an essential property of a price function, as stated in Definition~\ref{defn:arbitrage-freedom}.
In the query pricing strategy, there is no arbitrage in the pricing models of Q1, Q2, Q3, and Q5.
In addition, the higher social welfare is more important to the auction-based pricing model (i.e., Q4), and it is not certain whether there is arbitrage  in Q4.


\textbf{Market-oriented.} Whether to consider the market factors is an important factor to evaluate a data pricing model.  
In the query pricing strategy, almost all the pricing models are market-oriented.
For example, the studies \cite{chawla19revenue,koutris2013toward,miao2020towards,wang2019novel,zheng2017online,zheng2019arete}  take  market factors into account,  such as sellers' setting, buyers' valuation, revenue maximization, social utility, etc.
It means that, the query price is influenced by both the pricing mechanism and the whole data market.

\section{Feature-based Data Pricing}
\label{sec:fbmodel}


In this section, we elaborate four representative pricing models in the feature-based data pricing (FP) strategy, and then we compare them from several aspects.


The FP strategy focuses on establishing a price for a dataset, instead of pricing a query like the QP strategy described in Section~\ref{sec:qpmodel}.
The FP strategy determines the data price based on data features, including the pricing model  based on quality (i.e., F1), the personal data pricing model using privacy (i.e., F2), the personal data pricing model with market value (i.e., F3), and the social network pricing model with influence (i.e., F4). The underlying idea of the FP strategy is that, it is possible to derive its \emph{intrinsic} value that could be represented by data \emph{features}, even though it is hard to obtain the \emph{true} value of data.




\vspace*{-0.05in}
\subsection{Data Pricing based on  Quality}
\label{subsec:fb-quality}
\vspace*{-0.05in}

The data pricing model  based on quality (i.e., F1) attempts to price various data from the perspective of data quality in different dimensions.

The  price derived by F1  concentrates on the datasets' statistic property, and thus it is easily understood and employed.
The idea of  F1 has been widely used in real life.
For example, YoueData adopts the incompleteness degree as a factor of rating the sold data~\cite{youedata}.
However, the feature of data quality only partly reflects the inherent value of data, which is insufficient to represent the true  value of data.



First, based on the idea of ``what you pay for is what you get'', a theoretical and practical pricing framework trades data quality (more specifically, data accuracy) for discounted prices \cite{tang2013you}.
 In particular, the data buyer is free to propose his own price for data.
If the data consumer is able to afford the full price $p$ of the data, he/she will get them directly.
Otherwise,  an inaccurate version of data is returned to the data consumer according to his/her bid $p_b$ and the full price $p$.
The lower the price, the more inaccurate the answer.
 This model also belongs to posted pricing mechanism, where the price is posted by the buyers, instead of sellers/markets in previous pricing models.

To be more specific, the true value of each tuple is modeled as a degenerate distribution.
In the case of a discounted price, the inaccurate tuple value is randomly determined from a probability distribution.
The distance of the distribution to the degenerate distribution is measured through earth movers distance (EMD), which is commensurate to the discount, i.e., $p_b/p$.
Moreover, the returned tuple value is drawn from the distribution with a guaranteed probability,  which is also commensurate to the discount.
Each tuple in required data is treated through the above process, thus the whole inaccurate data are traded to the consumer.


Furthermore, following the idea of ``what you pay for is what you get'', a pricing framework for trading XML data considers the case of XML documents with completeness as the trading products \cite{tang2014get}.
The data completeness is traded for a discount.
Specifically, the XML data is formed as a tree, and the completeness is defined as the weight ratio between the full tree and answered tree.
A sampling problem is derived  to find a subtree that satisfies the weight, which is with respect to a discount price.
Meanwhile, the pricing mechanism is proved to be arbitrage-free, and the situation where consumers have repeated requests is also considered.

Moreover, a novel pricing framework considers multi-dimensional quality features~\cite{yu2017data}, instead of solely considering one single data quality feature.
It has the assumptions that, there are multiple versions of datasets, and prices are based on multi-dimension quality.
In particular, the data cost is modeled as a linearly increasing function of quality levels on different dimensions.
Also, the willingness of data consumers is represented through a piecewise function based on multi-dimensional quality levels.
As a result, a bi-level programming problem is formulated for revenue maximization, and a genetic algorithm is proposed to solve it.

\subsection{Personal Data Pricing using  Privacy}
\label{subsec:fp-privacy}

The amount of data in real-world is increasing at an amazing speed, and personal data accounts for a large part of it.
For instance, Facebook collected 300 petebytes of personal data in 2014~\cite{vagata2014scaling}.
However, personal data are difficult to share and use directly, due to the critical privacy concern.

The data market offers an important way to bridge the gap between sellers and buyers who are interested in personal data.
The personal data pricing model using privacy (i.e., F2) takes into consideration the growing public concern about data privacy to promote personal data circulation.





\begin{figure}[t]
	\centering
	\includegraphics[width=0.47\textwidth]{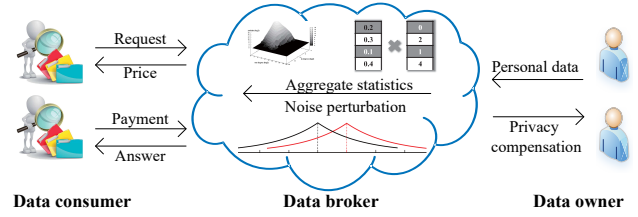}
	\vspace*{0.05in}
\caption{The framework of pricing private data \cite{li2014theory,li2017theory}}
	\label{fig:PData}
	\vspace*{0.05in}
\end{figure}

The problem of pricing private data is firstly studied in \cite{li2014theory,li2017theory}  with
a typical pricing framework based on data privacy  shown in Figure~\ref{fig:PData}.
Similar as existing data markets, the pricing framework includes three parts, i.e., data owner, data broker, and data consumer.
The data broker is an intermediary agent between data owners and data consumers, which deals with the request from data consumer, determines monetary compensation of data owner, prices the answer, and returns the noisy answer to data consumer.

In particular, the request from data consumer can be formalized as $S=(\mathbf{q}, v)$, where $\mathbf{q}$ can be seen as a function that outputs some aggregate statistics, such as linear aggregate query, histogram count, weighted sum, mean, standard deviation, etc.  The whole database formulated as a data vector $\mathbf{x}$ can be seen as the output of arbitrary function on the original data.
For example, each element $\mathbf{x}_i$ in $\mathbf{x}$ is the number of records in specified domains, e.g., ``is the age greater than 18''~\cite{li2017theory}.

Moreover, the request $\mathbf{q}$ is also a vector, and the true answer of $\mathbf{q}(\mathbf{x})$ can be formulated as the vector product, i.e., $\mathbf{q} \mathbf{x}=q_{1} x_{1}+\cdots+q_{n} x_{n}$.
Note that, the data broker does not return the \emph{raw} data (like traditional SQL query does), since the original data is sensitive.
The parameter $v$ is the tolerable variance of noise added to the true answer.
Once the data broker receives the request, he/she not only needs to return the noisy answer, but also has to derive the price and privacy compensation.


Based on this framework, two critical items in pricing problem are introduced, namely, (a) how to ensure privacy and compensate the data owners, and (b) how to derive \emph{arbitrage-free} price function.
To protect the privacy, \emph{Laplace} mechanism is employed to guarantee the dependent differential privacy~\cite{dwork2011firm}.
In other words, the returned answer is processed by adding noise to perturb the true value of answer, i.e., $S(\mathbf{x}) = \mathbf{q}(\mathbf{x})+Laplace(b),$ and $b = \sqrt{v / 2}$, so as to make it hard to infer private information.
By comparing the output of the randomized mechanism $\mathcal{M}$ over the data vector $\mathbf{x}$ and the data vector with absence of data owner $\mathbf{x}^{(i)}$ \cite{ghosh2011selling}, the individual privacy loss can be formulated as stated in Definition~\ref{defn:indi-privacy-loss}, which is used in~\cite{li2014theory,li2017theory,niu2018unlocking,niu2019erato}.


\begin{definition}\label{defn:indi-privacy-loss}
{\bf (Individual privacy loss)}.
The privacy loss of the data owner $i$ in the randomized mechanism $\mathcal{M}$ over the database $\mathbf{x}$ is defined as:
\begin{equation}
\epsilon_{i}(\mathcal{M})=\sup _{\mathbf{x}, O}\left|\log \frac{\Pr(\mathcal{M}(\mathbf{x})=O)}{\Pr\left(\mathcal{M}\left(\mathbf{x}^{(i)}\right)=O\right)}\right|
\end{equation}
\noindent
where $\mathbf{x}$ is the data vector, $\mathbf{x}^{(i)}$ simulates the absence of data owner $i$, and $O$ ranges over all possible outputs.
\end{definition}

In particular, the privacy loss measures the influence of data from owner $i$ on the output (denoted by a random variable $\mathcal{M}(\mathbf{x})$).
Furthermore, the upper bound of the privacy loss on data owner $i$ is found, as follows.
$$\epsilon_{i}(S) \leq \frac{\sup _{\mathbf{x}, i}\left|\mathbf{q}(\mathbf{x})-\mathbf{q}\left(\mathbf{x}^{(i)}\right)\right|}{\sqrt{v / 2}}$$

Thus, a payment for data owner $i$, i.e., the privacy compensation, is derived below.
$$\mu_{i}(S)=c_i\cdot\epsilon_{i}(S),\quad c_i \text{ is a constant},c_i>0$$
Given the payment for data owners, the whole price of $S$ is formulated in the following.
\begin{equation}
\label{eq:pdataprice}
    \pi(\mathbf{q}, v)=\pi(S)=\sum_{i} \mu_{i}(S)
\end{equation}

In order to ensure the arbitrage freeness,
the price functions $\pi(\mathbf{q}, v)$ following these properties are proved to be arbitrage-free~\cite{li2014theory,li2017theory}.

\begin{proposition}
\label{prop:arb-privacy}
Any arbitrage-free price function $\pi(\mathbf{q}, v)$ satisfies the following properties,  where $\mathbf{q}$ refers to data analysis method that produces output, and $v$ is the noise level.

\noindent
$(i)$ The zero $\mathbf{q}$ is free: $\pi(\mathbf{q}, v)=0$.

\noindent$(ii)$ Higher variance is cheaper: $v \leq v^{\prime}$ implies $\pi(\mathbf{q}, v) \geq \pi(\mathbf{q}, v^{\prime}) $.

\noindent$(iii)$ The zero-variance $\mathbf{q}$ is the most expensive: $\pi(\mathbf{q}, 0) \geq \pi(\mathbf{q}, v) \text { for all } v \geq 0$

\noindent$(iv)$ Infinite noise is free: if $\pi$ is continuous at $\mathbf{q}=0,$ then $\pi(\mathbf{q}, \infty)=0$.

\noindent$(v)$ As $v \rightarrow \infty, \pi(\mathbf{q}, v) = \Omega(1 / v)$.
\end{proposition}


This proposition is different from  Proposition~\ref{pro:free-qpricing}, as it considers the variable noise $v$.
The last property ensures that, the price function is arbitrage-free only if it decreases lower than 1/$v$ .
In such case, data consumers cannot purchase $(\mathbf{q}, v)$ with lower price by combining a series of $(\mathbf{q}, v_{i})$ when $\frac{1}{m} \sum_{i=1}^{m} v_{i} \leq v$, and thus arbitrage is avoided.
It is easy to find that, the price function $\pi(\mathbf{q}, v)$ in Eq.~\ref{eq:pdataprice} is arbitrage-free.

It is worthwhile to note that, there are several differences between a practical framework \textsc{Erato} \cite{niu2018unlocking,niu2019erato} and previous studies \cite{li2014theory,li2017theory}.
First, the $\epsilon$-dependent differential privacy is guaranteed in \textsc{Erato}, which results in the different forms of the upper bound on privacy loss and price function.
Second, there is another payment function in \textsc{Erato} for data owner $i$, i.e., $\mu_{i}(S)=b_{i} \cdot \tanh \left(c_{i}\cdot  \epsilon_{i}(S)\right)$, for constants  $b_i$, $c_i>0$, which bounds the payments of data owners. Third, a top-down design is considered in \textsc{Erato}, where the payment budget $B$ for all data owners is allocated to each owner based on the $\epsilon_{i}(S)$.
Last but not least, the evaluation results show that, \textsc{Erato}
improves the utility of aggregate statistics, guarantees arbitrage freeness, and compensates data owners in a fairer way than the classical differential privacy based approaches~\cite{li2014theory,li2017theory}.
Furthermore, there is a pricing framework  \textsc{Horae} for time-series private data~\cite{niu2019making}.
First, \textsc{Horae} employs \emph{Pufferfish} privacy to quantify privacy  loss  under temporal correlations.
In particular, the Pufferfish privacy can capture the correlations in different timestamps, and thus it is suitable for time-series data.
\textsc{Horae} compensates data owners with distinct privacy strategies in a satisfying way.
Besides, the theoretical analysis and experiments confirm that, \textsc{Horae} not only guarantees good profitability at the data broker, but also ensures arbitrage freeness against cunning data consumers.

\label{subsec:fb-analysis}
\begin{table*}[tbp]
  \setlength{\tabcolsep}{2.5pt}
  \centering
  \caption{Comparison of Feature-based Data Pricing Models}
    \begin{tabular}{|c|c|c|c|c|c|c|c|}
    \hline
    \textbf{Model} & \textbf{Reference} & \textbf{Perspective} & \textbf{Mechanism} & \textbf{Presetting} & \textbf{Technique}  & \textbf{Arbitrage-free} & \textbf{Market} \\
    \hline
     F1 & \cite{tang2014get,tang2013you,yu2017data}      & data quality &\makecell[c]{posted pricing} &basic price  & \makecell[c]{sampling, \\bi-level programming}  & uncertain   & $\checkmark$ \\
    \hline
     F2 &\cite{li2014theory,li2017theory,niu2019making,niu2018unlocking,niu2019erato} &  data privacy & posted pricing & N/A  & differential privacy & $\checkmark$    & $\times$ \\
    \hline
     F3 &\cite{niu2020online}     & market value & online pricing & feature & L$\ddot{\text{o}}$wner-John ellipsoid & uncertain   & $\checkmark$ \\
    \hline
    F4 &\cite{Zhu2020snpricing}     & node influence & posted pricing   &N/A & sampling & uncertain   & $\times$  \\
    \hline

    \end{tabular}%
    \label{tab:pricing-fbmodels}
\vspace*{0.06in}
\end{table*}

\subsection{Personal Data Pricing with Market Value}
\label{subsubsec:db-pmarket}

In addition to the pricing model concerning data privacy, there is another pricing model based on the market value of data (w.r.t. F3), which prices data in an online fashion.
The goal of this model  F3 is to minimize the regret during the ongoing transaction for maximizing the revenue, where the market value is modeled as a feature-based function.
Hence, the pricing model F3 is suitable to adjust the price based on the market feedback, so as to obtain the long-term gains.

Inspired by existing feature based pricing work \cite{cohen2020feature}, the market value of differentiated product can be formulated as a function, as stated in Definition~\ref{defn:market-value-goods}.

\begin{definition}\label{defn:market-value-goods}
{\bf (Market value of the product)}.
The market value $v_t$ of a product $t$  in the feature based pricing setting is defined as
\begin{equation}
v_t = f\left(\mathbf{x}_{t}\right)+\delta_{t}
\end{equation}
\noindent
where $\mathbf{x}_{t}$ is the feature vector of product, $f\left(\mathbf{x}_{t}\right)$ is the mapping function from feature vector $\mathbf{x}_{t}$ to the deterministic part in market value, and $\delta_{t}$ models the uncertainty in market value.
\end{definition}

The price of each product mainly relies on its features, and the uncertainty in markets is also considered.
Note that, the deterministic part $f\left(\mathbf{x}_{t}\right)$ is an arbitrary function, and can support different kinds of functions in both linear and non-linear forms.
When the price $p_t$ is lower or equal to the market value $v_t$, the consumer decides to buy the product, otherwise the transaction fails, resulting in zero profit.
To achieve more profit, the optimal price $p_{t}^{*}$ is the market value $v_t$.

Since the knowledge of market value is limited, the data broker cannot directly determine the market value $v_t$.
Instead, he/she can only passively receive each product request $\mathbf{q}_t$, and then post a price $p_t$.
When the posted price is no more than the market value, i.e., $p_{t} \leq v_{t}$, the data broker can earn a revenue.
Thus, a revenue maximization problem is identified and transformed to a regret minimization problem that is solved in an online way \cite{cohen2020feature}.
In particular, it approximates the polytope-shaped knowledge set with ellipsoid to provide a worst-case regret of $O\left(n^{2} \log T\right)$, where $n$ is the dimensionality of feature $\mathbf{x}_{t}$ and $T$ is the number of rounds of posting the price $p_t$. This  is essentially the pure version of pricing mechanism in \cite{niu2020online}.

Moreover, the personal data pricing model with market value in \cite{niu2020online}  models  $\mathbf{x}_{t}$  as the privacy features of a request $\mathbf{q}_t$ with the reserve price constraint.
The differentiated product can be seen as the request $\mathbf{q}_t$  of data consumer, i.e., the concrete statistic and tolerable level of noise.
In this setting, the features can be obtained by the privacy compensation and other techniques.
The analysis and experiments demonstrate the usefulness to set a proper reserve price, and the feasibility of this online pricing mechanism.

\subsection{Pricing Social Network with Influence}
\label{subsec:fb-influence}

The rapid development of online social network applications makes social network attract  much attention, and thus generates the huge amount of economic value.

The online advertising becomes an essential way to promote the product via the word of mouth  effects.
For example, the expenditure of advertisement on social media will be over \$50 billion by 2020, according to Fortune's claim~\cite{Reuters2016}.
The information can be propagated widely and rapidly through social networks, influencing users' decisions and behaviours in the diffusion process.


The model of pricing social network with influence (i.e., F4) is born in the context of viral marketing and online advertising.
It aims to derive the prices for a set of nodes (i.e., online celebrities/users) in social network  based on the expected influence \cite{Zhu2020snpricing}.
It not only presents a solution for online advertising, but also first involves the data pricing in social network, considering the information of network structure.

Specifically, the advertisers select a seed set $S$ from a node set $C$ and pay them in order to market the product and obtain more profit, where $C$ is provided by the online social network providers, corresponding to the online users that have ability or willingness to promote a product.
Intuitively, the price of any set of users should effectively reflect the expected influence spread that these users can generate in the information diffusion process.

Therefore, the divergence between the price and the influence spread can be characterized by $(c \cdot \sigma(S)$ $-$ $\sum_{s_{i} \in S} p_{i})^{2}$, where $c$ is a constant representing the expected revenue to derive from influencing a node.
$\sigma(S)$ is the influence spread of a node set $S$, which  is proportional to the probability that $S$ intersects with a random reverse reachable set \cite{borgs2014maximizing}. $p_i$ is the price of the node $s_i$. What's more, since the advertisers can choose any subset of nodes in $C$, the minimization problem of expected divergence between the price and influence spread  of any random chosen set is formulated as
$$
\min \quad \frac{1}{2^{|C|}} \sum_{S \subseteq C}\left(c \cdot \sigma(S)-\sum_{s_{i} \in S} p_{i}\right)^{2}$$
This problem is proved to be \#P-hard. There is an advanced algorithm to estimate the price profile with accuracy guarantees, which employs sampling technique to achieve a $(\varepsilon, \delta)$-approximation.

As a result, this pricing model finally aims to price nodes in a social network  (which is a special kind of data) via minimizing  the divergence between prices and influence of nodes. It is the first data pricing model in social networks, which is suitable for  viral marketing and online advertising.

\subsection{Analysis on Feature-based Data Pricing Models}
In this part, we compare the four representative feature-based data pricing models, as summarized in Table~\ref{tab:pricing-fbmodels}.

\textbf{Pricing perspective.}
The data quality based pricing model F1 trades low quality data with a lower price.
The pricing model using data privacy F2 derives the price based on the privacy loss of data owners.
The pricing model with market value F3 decides the price in an online fashion, making the price closer to data's true market value.
The pricing model for social network F4 assigns the price to a node  based on its influence.


\textbf{Pricing mechanism.}
The pricing models F1, F2, and F4 employ the posted pricing scheme, where the prices are posted by buyers (e.g., two models \cite{tang2014get,tang2013you} in F1) or sellers (e.g., F2 and F4).
While the pricing model F3 obtains the optimal price in an online fashion, and thus it belongs to the online pricing mechanism.

\textbf{Presetting.}
For the pricing model F1, there is a requirement for a predetermined price as the basic price of original data.
The pricing model F3 asks data features in  buyer's request before pricing, and the features should be carefully selected.
The other two pricing models, i.e., F2 and F4, generate the data prices in different ways, which do not need to preset some parameters.

\textbf{Techniques of price computation.}
The sampling technique is employed in F1, in order to derive a ``worse'' version of data, while the bi-level programming is used to tackle the revenue maximization problem in F1.
The differential privacy is used not only to protect the privacy of private data,  but also to quantify the privacy loss of data owners in F2.
As for the pricing model with market value F3, the L$\ddot{\text{o}}$wner-John ellipsoid method is helpful to support the price optimizing process.
Meanwhile, F4 utilizes the sampling technique  to boost the computation efficiency of nodes' influence spread.

\textbf{Arbitrage-free property.}
First, the pricing model with privacy  F2 and one pricing solution (i.e., data pricing based on completeness \cite{tang2014get}) in F1 consider the arbitrage freeness in the pricing framework.
In contrast, the other pricing models in F1, F3 and F4  do not take this property into consideration.
In other words, there is an opportunity for consumers to buy more information with a lower price.
For the online pricing model F3, it pays more attention on the market revenue, compared to the arbitrage issue.
The arbitrage-free property needs to be further explored in various real-life scenarios.

\textbf{Market-oriented.}
The data quality based pricing model F1 is market-oriented, since the willingness of buyers for versions of data is considered, and a revenue maximization problem is solved.
Meanwhile, the pricing model with market value F3 is obviously  market-oriented, since the optimal price is the market value.
The other two pricing models F2 and F4 are not market-oriented, since no market factor is considered in them.

\section{Pricing in Machine Learning}
\label{sec:mlmodel}

In this section, we introduce and compare five representative pricing models in the machine learning pricing (MLP) strategy. 

Data analytics using machine learning (ML) is an integral
part of science, business intelligence, journalism, and many other domains.
Research and industrial efforts have largely
focused on performance, scalability and integration of ML
with data management systems \cite{low2014graphlab,zhang2016materialization}.
As known, the large-scale and high-quality dataset is able to boost the performance of machine learning models~\cite{hestness2017deep}.

Data market is an efficient way to acquire quality guaranteed data for ML based data analytics.
It is important to note that, previous pricing schemes in data markets either force users to buy the whole dataset or support simplistic pricing mechanisms, without any awareness of the ML task downstream (e.g., the dataset is used to train a predictive model).  
Hence, users buy rich structured (relational) data to train their ML models, either directly through companies (e.g., Bloomberg \cite{bloomberg}, Twitter \cite{twitter}), or through data markets (e.g., BDEX \cite{bdex}, Qlik \cite{qilk}).
However, such datasets are often very expensive due to the immense effort that goes into collecting, integrating, and cleaning them.
It means that, valuable datasets may not be affordable to potential buyers with limited budgets, making data sellers operate in an inefficient market (without revenue maximization). 

\begin{figure}[t]
	\centering
	\includegraphics[width=0.4\textwidth]{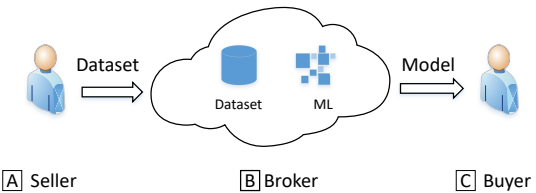}
	\caption{The market setup of pricing in MLP strategy \cite{chen2019towards,liu2020dealer}}
	\label{ML basics}
\end{figure}

Accordingly, the machine learning pricing (MLP) strategy becomes aware of the ML task downstream.
It sells ML models and data, instead of solely selling the data \cite{jordan2019artificial}.
As depicted in Figure~\ref{ML basics}, the data market selling ML models involves three agents, i.e., the seller who
provides the datasets, the buyer who is interested in buying
ML model instances, and the broker (market) who interacts
between the seller and the buyer.
The seller and/or the broker perform market research to ascertain curves representing demand and value for the ML model instances among potential buyers.

\begin{itemize}\setlength{\itemsep}{-\itemsep}
	\item{} The seller is the agent who wants to sell ML model instances
	trained on their profitable dataset $D$.
	\item{} The broker is the agent that mediates the sale for a set of supported ML models and gets a reward from the seller for each sale.
	\item{} The buyer is the agent interested in buying
	an ML model instance trained on $D$.
\end{itemize}


In the following, we elaborate four pricing models in MLP strategy, i.e., M1---the \emph{noise-injection} pricing mechanism to trade ML models, M2---the \emph{shapley value-based} data pricing scheme, M3---the \emph{end to end} pricing framework for selling both data and model, and M4---data pricing in \emph{federated learning}.


\subsection{The Noise-injection Pricing Mechanism}





When a buyer requests an ML model instance, the broker adds random \emph{Gaussian} noise to the optimal model and returns it to the buyer, so as to charge a price according to the variance of the noise injected.
It is the core idea of the noise-injection pricing mechanism, i.e., M1 \cite{chen2019towards}.

The pricing model M1 enables the buyer to either choose cheaper but less accurate instances or more accurate yet more expensive ones.
In other words, adding noise with low variance implies a model instance with expected low error, and thus a high price.
While noise with high variance results in an instance with expected larger error and a low price.
Essentially, the  pricing model M1 provides different versions of the desired ML model of varying quality, in analogy to the notion of versioning in information selling \cite{chen2019demonstration,shapiro1998versioning}.

Specifically, in this pricing mechanism, the \emph{seller} provides the dataset $D$ for sale, and it is given as a pair ($D_{\text{train}}$, $D_{\text{test}}$), where $D_{\text{train}}$ is the training set (for obtaining model instances) and $D_{\text{test}}$ is the test set.
The broker specifies a menu of ML models $\mathcal{M}$ that can support, along with the corresponding
hypothesis spaces $\mathcal{H}_m$ for each ML model $m \in \mathcal{M}$.
For now, fix an ML model, i.e., the hypothesis space $\mathcal{H}$. An error (loss) function $\lambda(h,D)$ measures the goodness of a hypothesis $h \in \mathcal{H}$ on $D_{\text{train}}$, and returns a real number in $[0, \infty)$.
Moreover, there is a group of hypothesis spaces $\mathcal{H}_m$ for each ML model $m$.
Given $D$ and the error function $\lambda$, let $h_{\lambda}^{*}(D)=\arg \min _{h \in \mathcal{H}} \lambda(h, D)$ denote the optimal model instance, i.e., the model instance that obtains
the smallest error on the training dataset w.r.t. $\lambda$.


In this  mechanism, the \emph{broker} releases a model instance through a randomized mechanism $\mathcal{K}$, which is enabled to trade off ML error for the price that the model instance is sold for. Specifically, $\mathcal{K}$ uses a set of parametrized probability distributions $\left\{\mathcal{W}_{\delta} | \delta \in \mathbb{R}_{+}\right\}$.
Given a dataset $D$, an error function $\lambda$ and a noise control parameter (NCP) $\delta$, the broker first computes the optimal model instance $h^*_\lambda(D)$. Then, he/she samples $w \sim \mathcal{W}_{\delta}$ and outputs a noisy version of the optimal model, $\hat{h}_{\lambda}^{\delta}(D)=\mathcal{K}\left(h_{\lambda}^{*}(D), w\right)$. The NCP $\delta$ will be used as a knob to control the amount of noise added, and in turn, the price
of the model instance sold.


\vspace*{0.01in}
On the other hand, the \emph{buyer} specifies an interested ML model $m \in \mathcal{M} $ learning over $D$, along with his/her preferences for the error function to use from the ones the broker supports. After a set of \emph{interactions} with the broker, the buyer obtains an instance of $m$ that satisfies his/her price and/or error constraints.
The interactions between the seller and broker, as well as between the broker and the buyer run as follows.


\vspace*{0.035in}
\textbf{Broker-seller interaction}.
Apart from providing $D$, the seller works with the broker to determine the price function $p$ for a given ML model.
The price function does not depend solely on the released model instance $\hat{h}_{\lambda}^{\delta}(D)$. Instead, it depends on $D$, the NCP $\delta$, and the error functions. In the interaction, the broker is able to set the price functions based on two curves provided by the seller  based on his/her market research about $D$.
These curves tell the  broker  how much value potential customers attach to model errors in terms of monetary worth (w.r.t. the value curve) and how much demand there is in the market for different model errors (w.r.t. the demand curve). Finally, the arbitrage-free prices are ensured through the monotone and subadditive constraints, and derived to maximize the revenue.

\vspace*{0.035in}
\textbf{Broker-buyer interaction}. The buyer-broker interaction has four steps.
(i) First, the buyer  specifies the ML model  that he/she is interested in $\mathcal{H}$ and the error function corresponding to that model.
(ii) Second, the  broker computes a curve that plots the price together with the expected error for every NCP $\delta$.
This curve shows to the buyer the possible price points of the different versions of this model.
(iii) At the third step, the buyer has three options. First, he/she can specify a particular point on the curve (i.e., a price-error combination). Since $\delta$ behaves monotonically
w.r.t. the expected error, the broker can find the unique $\delta^*$
that corresponds to that point, and obtains $\hat{h}_{\lambda}^{\delta^{*}}(D)$.
The second option is that the buyer specifies an error budget.
The third and final option for the buyer is to specify a price budget to the broker.
In the latter two options, the broker has to finally solve an optimization problem, in order to derive the ``best'' model instance under the error/price budget.
Then, the buyer pays the price $p$ to the broker.
Note that, the whole pricing scheme still belongs to the posted pricing mechanism, since the price is either posted by the buyer (in option 3) or the broker (in option 1 and 2).
(iv) Finally, the broker gives the obtained model instance $\hat{h}_{\lambda}^{\delta^{*}}(D)$ to the buyer.

\subsection{Pricing with Shapley Value}


The \emph{shapley value} (SV) coincides with people's intuition of data value. For instance, noisy images tend to have lower SVs than the high-fidelity ones.
The training data whose distribution is closer to the test data distribution tend to have higher SVs. Intuitively, the SV is the relative value of data which measures the marginal improvement of utility attributed to the data point, averaged over all possible subsets of data points.

As a consequence, in the pricing model with shapley value (i.e., M2), the concept of the shapley value is adopted to price the general machine learning models \cite{ghorbani2019data} and the family of ML models relying on $k$-nearest neighbors (KNN)~\cite{jia2019efficient}, both of which consider two types of agents that interact in a data market: the seller (or data curator) and the buyer.
The seller provides training data instances, each of which is a pair of a feature vector and the corresponding label.
The buyer is interested in analyzing the training dataset aggregated from various sellers and producing an ML model.
The goal is to distribute the payment given by the buyer fairly between the sellers.
A natural way to tackle the question of revenue allocation is to view ML as a cooperative game and model each seller as a player.

This game-theoretic viewpoint allows to formally characterize the ``power'' of each seller and in turn determine their deserved share of the revenue, as stated in Definition~\ref{defn:shapley value}.

\begin{definition}\label{defn:shapley value}
{\bf (Shapley value)}.
Given a utility function $U(\cdot)$ and a dataset $D$  with the size of $N$, the shapley value of the datum $i$  is defined as Eq.~\ref{eq:sv-1} or Eq.~\ref{eq:sv-2}.
\begin{equation}\label{eq:sv-1}
sv_{i}=\sum_{S \subseteq D \backslash\{i\}} \frac{1}{N\left(\begin{array}{l}
N-1 \\
|S|
\end{array}\right)}[U(S \cup\{i\})-U(S)]
\end{equation}
\begin{equation}\label{eq:sv-2}
  sv_{i}=\frac{1}{N !} \sum_{\pi \in \Pi(D')}\left[U\left(P_{i}^{\pi} \cup\{i\}\right)-U\left(P_{i}^{\pi}\right)\right]
\end{equation}
\end{definition}

The utility function $U(\cdot)$ is the evaluation of dataset on a specific model, such as the accuracy of $k$-nearest neighbors and random forest regression.
One can find that, the shapley value in Eq.~\ref{eq:sv-1} is defined as the average marginal contribution of datum $i$ to all possible subsets of the whole dataset.
In Eq.~\ref{eq:sv-2}, $D'$ is any possible subset on $D$ without datum $i$, $\pi \in \Pi(D')$ is a permutation of data $D'$, and $P_{i}^{\pi}$ is the set of data preceding datum $i$ in $\pi$.
The central idea behind Eq.~\ref{eq:sv-2}  is to regard the \emph{shapley value} definition as the expectation of a training instance's marginal contribution over a random permutation and then use the sample mean to approximate it.
Meanwhile, this two definitions are equal.
As \emph{shapley value} can quantify the contribution of data in machine learning models, it is a way to price data in machine learning fairly.
However, it is not a market-oriented pricing model, as it measures the influence of data in machine learning models, without considering any market factor.
The pricing model M2 can be extended to adapt to the market by combining market factors and valuation functions $U(\cdot)$.

\vspace*{0.1in}
\subsection{End to End Pricing Framework}


As described, the noise-injection pricing mechanism M1 focuses on pricing a set of model instances depending on the model quality to maximize the revenue.
While the pricing model with shapley value M2 considers how to allocate compensation in a fair way among data owners when their data are utilized for the machine learning models.
Different from prior pricing models M1 and M2 which either sell ML models or trade training data samples, there is a novel pricing model in MLP strategy, i.e., an end to end pricing framework for both ML models and data used in machine learning.

The end to end pricing model (i.e., F3) not only takes the compensation of data owners into account, but also solves the revenue maximization problem for model pricing.
The noise injection mechanism is also employed to train model with noise based on differential \emph{privacy}.
Meanwhile, the pricing framework is a market-oriented model, since it considers the market factors and solves the revenue maximization problem.


\textsc{Dealer}~\cite{liu2020dealer} is a typical end to end pricing framework.
It acts as a broker between model buyers and data owners. The whole dynamic process of transaction in \textsc{Dealer} can be formulated as follows.

\begin{itemize}
  \item \textbf{Data collection.} It collects  a dataset $D=\{D_{1},D_{2},$ $\cdots,D_{n}\}$, and computes the compensation $c_{i}(\epsilon)$ of each datum $D_i$ based on shapley value and differential privacy.
  \item \textbf{Model parameter setting.} \textsc{Dealer} decides a set of ML models to train with different coverage rates and differential privacy parameters.
  \item \textbf{Model pricing.}
  \textsc{Dealer} conducts a market survey among $m$ sampled model buyers, and collects results on model demand and corresponding valuation of each model. Then, it computes the optimal price $p_k$  for each ML model $M_{k}$  to maximize the revenue.
  \item \textbf{Model training and release.}
  \textsc{Dealer} first selects the training subset  $\mathcal{S}_{k}$ with budget constraint to maximize the shapley value of $\mathcal{S}_{k}$.
  Then, \emph{Dealer} trains the model $M_{k}$ using  $\mathcal{S}_{k}$, and guarantees coverage rate $\mu_{k}$ and differential privacy parameter $\epsilon_{k}$ in model parameter setting step.
  \item \textbf{Model transaction.} The model buyer pays $p_i$ for his target model $M_i$ when his valuation $v_i\ge p_i$,
  and the broker sends the corresponding model to the buyer.
  \item \textbf{Compensation allocation.} \textsc{Dealer} allocates the corresponding compensation to $D_{i}$ based on $c_{i}(\epsilon)$ if $D_i$ is used to train model.
\end{itemize}

In particular, there are two main problems during the process of \textsc{Dealer}.
One is how to derive models' prices for revenue maximization without arbitrage.
Another is how to choose the training subset $\mathcal{S}_{k}$  from dataset for an ML model $M_{k}$ with the budget constraint, and allocate  the compensation for datum in  $\mathcal{S}_{k}$.
For the revenue maximization problem, the buyers' valuation on machine learning models is defined as a function of the coverage rate $\mu_{k}$ and privacy rate $\epsilon_{k}$.
Then, the subadditivity constraint of price functions (for arbitrage-free concern) is relaxed.
It derives the prices of models using an efficient dynamic programming algorithm.
Furthermore, in order to select the subset $\mathcal{S}^{k}$ quickly, a data coverage maximization problem is addressed with pseudo-polynomial time algorithm and enumerative guess-based greedy algorithm.

\begin{table*}[tbp]
  \setlength{\tabcolsep}{2pt}
  \centering
  \caption{Comparison of Pricing Models in Machine Learning}
    \begin{tabular}{|c|c|c|c|c|c|c|c|}
    \hline
    \textbf{Model} & \textbf{Reference} & \textbf{Perspective} & \textbf{Mechanism} & \textbf{Presetting} & \textbf{Technique}  & \textbf{Arbitrage-free} & \textbf{Market} \\
    \hline
      M1  & \cite{chen2019towards}    &  model quality & posted pricing &market survey  & noise injection  & $\checkmark$    & $\checkmark$ \\
    \hline
      M2  & \cite{ghorbani2019data,jia2019efficient}     & data contribution & posted pricing & evaluation & approximate technique & uncertain    & $\times$ \\
    \hline
      M3  & \cite{liu2020dealer}    &  \makecell[c]{differential privacy,\\shapley value} & posted pricing & market survey  & \makecell[c]{dynamic programming,\\ differential privacy, etc.} & $\checkmark$   & $\checkmark$ \\
    \hline
       M4  & \cite{bao2019flchain,feng2019joint,hu2020trading,jiao2020toward}     & customer incentive & \makecell[c]{ posted pricing, \\auction} & N/A  & federated learning & uncertain    & $\checkmark$ \\
    \hline

    \end{tabular}%
    \label{tab:pricing-mlmodels}
    \vspace*{0.09in}
\end{table*}%

\vspace*{0.1in}
\subsection{Pricing in Federated Learning}


Federated learning (FL) is introduced by Google~\cite{mcmahan2017communication}, which distributes the learning process to individuals and then collaboratively trains a target model.

The model trained in FL framework keeps the data on users' devices that does not need to collect all the data together, and thus alleviates the privacy issue.
FL is apparently superior to traditional data analysis and machine learning techniques, that require all the data to be collected to a centralized data center or server for producing  effective machine learning models (which may raise data security and privacy issues).

The pricing model in federated learning (i.e., F4) is indispensable to promote the participation of a large base of data owners, so as to build a healthy federated learning service market \cite{bao2019flchain,feng2019joint,hu2020trading,jiao2020toward}.
For example, FLChain~\cite{bao2019flchain} is a decentralized, public audible, and healthy federated learning ecosystem with trust and incentive.
It evaluates reliability and contribution of data owners and offers fair price for them.
Thus, FLChain is able to motivate trainer to be honest and detect  report misbehavior, which can provide a healthy market for collaborative-training models.

There are so far two pricing schemes encouraging the clients to participate the federated learning~\cite{feng2019joint,jiao2020toward}.
In particular,  the price in~\cite{feng2019joint} is  related to the training data size.
Considering data pricing and cooperate relaying jointly, it utilizes a Stackelberg game model to analyze the transmission strategy, data pricing strategy, and service subscription in the FL system.
On the other hand, an auction based market model is proposed in~\cite{jiao2020toward} to incentivize data owners to participate in federated learning, where the utilities of each data owner and the federated learning platform are both considered.
It finally formulates a social welfare maximization problem.
This problem is solved via a reverse multi-dimensional auction mechanism that decomposes the original auction mechanism into a set of sub-auctions.
The solution is proved to be truthful, individually rational, and computationally efficient.



Furthermore, a new pricing mechanism for federate learning is proposed, which considers the privacy of data \textcolor[rgb]{0.00,0.00,0.00}{owners}, allows data owners to choose a privacy budget, and offers price according to the privacy loss~\cite{hu2020trading}.
The training process is combined with privacy concerns, and the utility of FL server is formulated as difference between the influence of noise (for privacy) on model accuracy and the prices for all used data.
As a result, the whole mechanism is modeled as a two-stage Stackelberg game.
It is necessary to mention that, the former pricing schemes~\cite{bao2019flchain,feng2019joint,hu2020trading} do not consider market factors.
In contrast, the pricing model in \cite{jiao2020toward} is market-oriented, as it considers the utility of both data owners and federated learning platforms.

\subsection{Analysis on Pricing in Machine Learning}
\label{subsec:ml-sum}

We compare the four pricing models in MLP strategy from several aspects, as summarized in
Table \ref{tab:pricing-mlmodels}.

\textbf{Pricing perspective.}
The noise-injection pricing mechanism M1 considers the model quality. It trades \emph{models} with different error levels.
The pricing model with shapley value M2 is a popular way to measure the contribution of each \emph{data sample}  in machine learning model training.
While the end to end data pricing framework M3 trains different versions of models, which are based on different subsets of data and privacy guarantees.
The subset of data is selected according to the coverage rate of its shapley value.
The compensation allocation on data is related to shapley value and differential privacy.
It means that, the whole pricing framework M3 prices \emph{data} and \emph{ML models} based on differential privacy and shapley value.
In addition, the pricing model in federated learning M4 prices data to incentivize data share  based on data reliability, transmission strategy, welfare maximization, or privacy concerns.

\textbf{Pricing mechanism.}
Most pricing mechanisms belong to posted pricing, such as the noise-injection mechanism and the end to end pricing framework.
While the pricing solution in federated learning \cite{jiao2020toward} utilizes auction to maximize social welfare.

\textbf{Presetting.} For noise-injection pricing mechanism M1 and the end to end pricing framework M3, they both require the market survey to derive better price, i.e., preparing to estimate the buyers' valuation, for maximizing the revenue.
Meanwhile, the pricing model with shapley value M2 requires a reasonable and good evaluation function to quantify data contribution.
There is no presetting requirement for the pricing model in federated learning M4.

\textbf{Techniques of price computation.}
In the MLP strategy, the noise-injection pricing mechanism M1 trains different versions of models by adding noise.
For M2, researchers explore approximate technique to efficiently compute the shapley value.
In M4, the pricing issues are related to the learning mechanism, and it requires related techniques to model the data cost, in order to solve the maximization problem.
In addition, there are several techniques employed in M3, such as dynamic programming, differential privacy, and noise injection.

\textbf{Arbitrage-free property.} There is no arbitrage in the noise-injection pricing mechanism M1 and the end to end pricing framework M3, i.e., they are arbitrage-free.
However, the pricing models M2 and M4 may suffer arbitrage, since they do not consider this property.

\textbf{Market-oriented.} The noise-injection pricing mechanism M1 and the end to end pricing framework M3 are market-oriented, as they need the market survey for preparation, in order to maximize the revenue.
The pricing model with shapley value M2 only considers the data contribution to ML models.
In addition, the goal of pricing model \cite{jiao2020toward} in M4 is to maximize the social welfare, and the utilities of both data owners and platform are considered, thus M4 is market-oriented.

\section{Future Directions}
\label{sec:future}

In this section, we identify a series of general research challenges in data pricing, and then we put forward several interesting topics for future directions.

\vspace*{-0.05in}
\subsection{General Challenges}
\label{subsec:challenges}


First, a data pricing model should be \emph{efficient} enough to widely apply to various data markets or systems.
Existing data pricing models basically have good theoretical properties, while they are not easy to be implemented in real data markets due to the high complexity of pricing algorithms.
For example, in the query pricing model with data view (i.e., Q1), the price derivation algorithms are NP-hard~\cite{koutris2012querymarket,koutris2013toward}.

\vspace*{0.02in}
Second, a data pricing model should be \emph{universal} to support a variety of data pricing scenarios.
Most of current data pricing solutions are developed to solve one kind of pricing problems under certain circumstances.
As an example, in the feature-based pricing (FP) strategy, most pricing solutions take one feature into consideration.
Taking multi-features into account is difficult for data pricing, since these features are correlated and may influence each other.
Moreover, the design of features relies on human experience, thus automatic feature extraction for data pricing is urgently required.
Thus, one has to make more efforts to achieve effective pricing methods applicable to more real-life scenarios.

\vspace*{0.02in}
Third, a data pricing model should be \emph{scalable}  for practical data trade over markets.
On the one hand, how to price the large-scale dataset reasonably is still a problem.
In this case, the dataset cost for collection, integration, and cleaning cannot be ignored, while existing methods do not consider it.
What's more, the price of large-scale dataset may influence the demand of small datasets, and the data markets should consider this issue and price data properly.
On the other hand, faced with numerous requests, data pricing algorithms should be able to process online price assignment tasks.
Therefore, the scalable data pricing model is beneficial for both large-scale dataset and massive requests.

\vspace*{0.02in}
Fourth, a data pricing model should be \emph{interdisciplinary} due to the diverse origins of pricing problem.
The pricing issue involves multiple disciplines, including microeconomics, operation research, management science, computer science, etc.
It finally comes to a series of totally different pricing solutions after the data pricing problem is studied in various backgrounds and fields.
However, there are huge gaps among the pricing methods from different disciplines.
Although there are some pricing mechanisms involving revenue maximization and social welfare maximization, it is hard to combine the advantages of these pricing methods (due to the different definitions of revenue and utility).
Moreover, how  to evaluate the data pricing solutions through interdisciplinary criteria is still open.
When deriving data prices under different pricing settings, it is also hard to conclude which pricing model is better.
Thus, it is necessary to utilize multi-disciplinary knowledge to further evaluate pricing methods fairly and reasonably.

\vspace*{0.02in}
Fifth, a data pricing model should offer an \emph{adaptive} mechanism to support the market dynamics and data update.
Most of existing pricing models only employ one pricing mechanism, such as posted  pricing, auction, etc.
However, data pricing is not for the one-off sale, and the changing value of data calls for \emph{adaptive} pricing schemes. 
Thus, different pricing mechanisms should be employed in different periods, in order to adapt to the dynamic changes of the market.
Also, the pricing mechanism is required to support the data update, and the price of data may be adjusted.

In the following, we provide some preliminary ideas on how to approach these challenges in principle.


\begin{itemize}\setlength{\itemsep}{-\itemsep}
\item \textbf{The efficiency}. It is expected that, pricing mechanisms with high efficiency are preferentially selected in data trade.
    First, data trade with blockchain techniques  is a promising way to achieve efficiency (and further address the privacy issue) \cite{xiong2019smart}.
    In addition, multi-granularity price options with different time constraints and approximate price computation can further enhance efficiency.

\item \textbf{The universality}.
    The universality is a profound property for a pricing model.
    First, it is an innovative attempt to establish pricing solutions from raw data to more complex data products, e.g., knowledge. Second, there is an urgent need for a universal standard for data pricing, which points out the necessary criteria, for example, arbitrage-free.
    In the gradual formation of the standard, many concerns will be solved, such as the ownership of data, the privacy protecting law, and so on. 

\item \textbf{The scalability}.	
To handle the large-scale dataset, it is natural to explore the cloud data pricing issue, e.g.,
how a cloud data service provider should activate and price optimizations that benefit many users~\cite{upadhyaya2012price}.
Moreover, the online and distributed implementations of existing pricing models are required to deal with the massive requests of buyers.

\item \textbf{The interdisciplines}.
The query pricing (QP) strategy and feature-based pricing (FP) strategy almost originate from the field of data management.
At the same time, the machine learning pricing strategy (MLP) is investigated in the machine learning community.
One can easily observe that, all of them have  more or less  integrated some fundamental theories or models from other disciplines, such as economics, management science, operation research, etc.
For instance, it is promising to leverage the game theory and auction models  to fill the interdisciplinary gaps between pricing and marketing.
Hence, the interdisciplines attempt to price data is just the beginning.

\item\textbf{The adaptability}.
On top of pros and cons of existing pricing mechanism, one can design a new pricing mechanism to automatically adjust in the dynamic markets.
The online pricing mechanism is somehow adaptive, such as Q5 and F3.
But it assumes that, the optimal price does not change in the whole process, which is impractical.
Moreover, for the data update, the stale data might be off the shelves, or depreciate like physical goods.
Some new data are probably similar to the stale data, which may lead the arbitrage issue.
Hence, it is necessary to present an adaptive pricing mechanism to adapt the dynamics in data and markets.
\end{itemize}


\subsection{Interesting Topics}
\label{subsec:topics}

In this part, we put forward several promising research topics related to data pricing.

\subsubsection{Data Acquisition}
\label{subsubsec:topic-acq}

Data acquisition, i.e., the process of sampling data from real world conditions, is of great importance for data analytics.
For example, modern machine learning models require  sufficient high-quality data for training.

Most data buyers directly obtain the dataset without caring about the underlying process, such as data collection, data integration, data cleaning, etc.
There exist two issues in such case, i.e.,  the process of acquiring data is not considered for data sellers in data markets, and it lacks on-demand data acquisition (discovery) service for data buyers.

First, regarding the problem of how to acquire data with the purpose of maximizing the profit of data markets or sellers, the researchers in Q5 to some extent targets such a problem in mobile crowd sensing~\cite{zheng2017trading}.
However, there is lack of more effective pricing solutions to incorporate data acquisition.
For example, the crowdsourcing technique is a widely-used way in collecting data~\cite{yang2020game}.
It is worthy to study of how to collect data via crowdsourcing based on the demand of data markets.
Moreover, the price of both data and crowdsourcing tasks can be considered in an end to end way.

Second, how to meet various data needs of data buyers has also not been fully explored.
It is common that, the dataset is (partially) ready in data markets, while data buyers cannot directly find what they desire.
Hence, the techniques of dataset search~\cite{brickley2019google,chapman2020dataset}, data recommendation, and data rebuild~\cite{fernandez2020data,li2018cost} are attractive in such scenarios.
Up to now, there is no prior work on dataset search in data markets.
Meanwhile, there is lack of in-depth explanations and practical solutions to the problems of data recommendation and data rebuild in data markets.
Therefore, the data acquisition (discovery) problem to help buyers find their desirable data is urgently needed in the modern data markets.

\subsubsection{Data Valuation}
\label{subsubsec:topic-value}

In addition to data acquisition, data valuation is equally important to data share and analytics.

On the one hand, assessing the value of datasets is helpful to data pricing and data trade \cite{heckman2015pricing}.
On the other hand, data valuation benefits the model training in several aspects, which facilitates the development of artificial intelligence (AI) techniques \cite{koh2017understanding,koh2019accuracy,yoon2019data}.
These solutions are useful to quantify the specific role of data in model training, leading to data pricing and machine learning models more explainable.

There are more interesting topics worth studying in terms of data valuation in data markets.
First, existing literatures focus on the quality of data, instead of the value of data, since the value of data is hard to reflect.
It is much meaningful to evaluate the valuation for the whole dataset, which could directly provide the basis for data pricing.
Meanwhile, rethinking the data value in certain scenarios, such as recommendation system and online advertising, may be useful.
For example, data pricing in social networks (i.e., F4) is born in the context of online advertising, and the prices are based on the data's propagation ability.
It is an effective way to connect data valuation with real-life applications, so as to make the data value more clear and convincing.

At the second place, most of existing studies aim to quantify the contribution of the datum and a group of data in machine learning and deep learning models  \cite{koh2017understanding,koh2019accuracy,yoon2019data}.
They are helpful to the AI field, but time consuming to data pricing.
Moreover, it is necessary to measure data contribution in certain models, like $k$NN classifier \cite{jia2019efficient}. 
Meanwhile, it is worthwhile to note that, although the work \cite{yoon2019data} enjoys more accuracy than existing studies \cite{ghorbani2019data,koh2017understanding}, it suffers the dilemma without theoretical guarantee.

Third, it is a promising approach to integrate data valuation with economic theory~\cite{glazer1993measuring}, e.g., answering the questions like how much profit the data bring, or how much cost the data reduce.
The economic value of data is highly related to its information value, and thereby a more appropriate market model for data goods is urgently needed, in order to better understand data value.
Also, the study on this problem is beneficial for the formation and evaluation of data assets, making data-driven industries more prosperous.



\subsubsection{Various Pricing Issues}
\label{subsubsec:topic-FL}
\vspace*{-0.02in}


\textbf{Diverse trading subjects.} 
The data become more diverse with the popularity of smart devices. 
For example, the current AI techniques are utilizing the multi-model data to train robust models~\cite{shen2019meal,zhang2020multi}.
Thus, the pricing mechanism that supports to price multi-type data is better than the one only for a specific type of data.
At the same time, the correlation between different types of data and prices should be rethinking.
Moreover, the more general concept of data pricing is apt to cover the pricing of sophisticated data services, personalized  data softwares, and so on. Existing data-based services adopt fixed or manual pricing methods, and thereby the smart pricing for various trading subjects is urgently needed.

\vspace*{0.02in}
\textbf{History-aware pricing.}
It is an economic solution for a data market to employ the history-aware pricing mechanism, i.e., the data buyers do not need to pay the data that they paid before.
This problem is currently solved in two ways, one is to enable the data pricing mechanism to be history-aware~\cite{deep2017qirana,miao2020towards}, while another is to employ additional mechanism to avoid the repeat payment via refunds~\cite{upadhyaya2015managing,upadhyaya2016price}.
However, how to establish an efficient and practical history-aware pricing mechanism is still open.
It has to address some issues like how to reflect the complex depreciation rules of different data assets into history-aware pricing models, how to efficiently identify the previously purchased data samples, and how to derive the history-aware price satisfying users' various constraints.

\vspace*{0.02in}
\textbf{Pricing in various platforms.}
To begin with, the federated learning (FL) model pricing is not taken into consideration in existing studies~\cite{bao2019flchain,feng2019joint,hu2020trading,jiao2020toward}, which assume the model marketplace is mature.
The FL model is different from traditional ML models.
For example, the data is distributed in users' devices, which makes the multi-version models require additional cost.
Moreover, existing model pricing schemes do not take the training cost into account. It is impractical.
Meanwhile, how to select a group of users for FL tasks is a problem, which requires the prediction on the quality, reliability, and cost of data.
Also, it is helpful to motivate more users to share data if establishing differentiated pricing between FL participants and external personnel.



\vspace*{0.02in}
On the other hand, as known, big companies have all built empires atop the data services, like Google cloud platform \cite{google}, Microsoft Azure Omnia \cite{omnia}, IBM dynamic pricing \cite{ibm}, and Amazon elastic compute cloud (Amazon EC2) \cite{Amazon,kumar2018survey}.
Although it seems like these companies are currently data (services) oligopolies, it is expected that, in the near future, there will appear more fair competition markets for the trade and share of data (services).
In addition, there are sufficient data that record data services' features and corresponding prices. It is helpful to derive proper prices from existing data and to evaluate pricing mechanisms.




\vspace*{0.02in}
\textbf{Data platform design.}
There is an urgent need for a practical data market, which allows researchers to validate  their data pricing studies in the platform.
Take the crowdsourcing platform Amazon Mechanical Turk (AMT) as an example, it not only performs as a financial platform to obtain profit via crowdsourcing, but also provides opportunities for scholars to conduct experiments in real world.
There are many concerns to conduct such a platform of data pricing.
For example, how to ensure the privacy of data and users, how to make sure that the data have not been pirated, how to determine the data ownership in various cases, etc.

\section{Conclusions}
\label{sec:conclusion}


In this paper, we carry out a comprehensive survey of data pricing models.
We present a hierarchical taxonomy of modern data pricing solutions with three pricing strategies, including query pricing QP, feature-based data pricing FP, and pricing in machine learning MLP.
We elaborate thirteen pricing models with in-depth analyses and comparisons.
We put forward five research challenges and a series of insights over data pricing.
Based on the survey study, we conclude that, data pricing is in its infancy, requiring more practical pricing schemes for various scenarios.

\balance

\begin{acknowledgements}
This work was supported in part by the NSFC Grants No. 62025206, 61902343, 61972338, 61825205, and 61772459.
Yunjun Gao is the corresponding author of the work.
\end{acknowledgements}

\balance
{
\bibliographystyle{spmpsci}
\bibliography{survey}

\begin{thebibliography}{100}
\providecommand{\url}[1]{{#1}}
\providecommand{\urlprefix}{URL }
\expandafter\ifx\csname urlstyle\endcsname\relax
  \providecommand{\doi}[1]{DOI~\discretionary{}{}{}#1}\else
  \providecommand{\doi}{DOI~\discretionary{}{}{}\begingroup
  \urlstyle{rm}\Url}\fi

\bibitem{AmazonProduct}
Amazon: https://www.amazon.com/

\bibitem{Amazon}
AmazonSpotInstances: https://docs.aws.amazon.com

\bibitem{armstrong1998mixing}
Armstrong, A.A., Durfee, E.H.: Mixing and memory: {E}mergent cooperation in an
  information marketplace.
\newblock In: ICMAS, pp. 34--41. IEEE (1998)

\bibitem{bao2019flchain}
Bao, X., Su, C., Xiong, Y., Huang, W., Hu, Y.: {FL}chain: A blockchain for
  auditable federated learning with trust and incentive.
\newblock In: BIGCOM, pp. 151--159. IEEE (2019)

\bibitem{bdex}
BDEX: https://www.bdex.com/

\bibitem{bloomberg}
Bloomberg: https://www.bloomberg.net/

\bibitem{borgs2014maximizing}
Borgs, C., Brautbar, M., Chayes, J., Lucier, B.: Maximizing social influence in
  nearly optimal time.
\newblock In: SODA, pp. 946--957. SIAM (2014)

\bibitem{brickley2019google}
Brickley, D., Burgess, M., Noy, N.: Google dataset search: Building a search
  engine for datasets in an open web ecosystem.
\newblock In: WWW, pp. 1365--1375. ACM (2019)

\bibitem{chapman2020dataset}
Chapman, A., Simperl, E., Koesten, L., Konstantinidis, G., Ib{\'a}{\~n}ez,
  L.D., Kacprzak, E., Groth, P.: Dataset search: {A} survey.
\newblock VLDB J. \textbf{29}(1), 251--272 (2020)

\bibitem{chawla19revenue}
Chawla, S., Deep, S., Koutris, P., Teng, Y.: Revenue maximization for query
  pricing.
\newblock PVLDB \textbf{13}(1), 1–14 (2019)

\bibitem{chen2019towards}
Chen, L., Koutris, P., Kumar, A.: Towards model-based pricing for machine
  learning in a data marketplace.
\newblock In: SIGMOD, pp. 1535--1552. ACM (2019)

\bibitem{chen2019demonstration}
Chen, L., Wang, H., Chen, L., Koutris, P., Kumar, A.: Demonstration of nimbus:
  Model-based pricing for machine learning in a data marketplace.
\newblock In: SIGMOD, pp. 1885--1888. ACM (2019)

\bibitem{cohen2020feature}
Cohen, M.C., Lobel, I., Paes~Leme, R.: Feature-based dynamic pricing.
\newblock Manag. Sci. \textbf{66}(11), 4921--4943 (2020)

\bibitem{cui2000practical}
Cui, Y., Widom, J.: Practical lineage tracing in data warehouses.
\newblock In: ICDE, pp. 367--378. IEEE (2000)

\bibitem{cui2000tracing}
Cui, Y., Widom, J., Wiener, J.L.: Tracing the lineage of view data in a
  warehousing environment.
\newblock ACM Trans. Database Syst. \textbf{25}(2), 179--227 (2000)

\bibitem{datagov}
Data.gov: https://www.data.gov/

\bibitem{datasift}
Datasift: https://datasift.com/

\bibitem{datasociety}
DataSociety: https://datasociety.net/

\bibitem{deep2016design}
Deep, S., Koutris, P.: The design of arbitrage-free data pricing schemes.
\newblock In: ICDT, pp. 12:1--12:18. Schloss Dagstuhl-Leibniz-Zentrum fuer
  Informatik (2017)

\bibitem{deep2017qirana}
Deep, S., Koutris, P.: {QIRANA}: A framework for scalable query pricing.
\newblock In: SIGMOD, pp. 699--713. ACM (2017)

\bibitem{deep2017qirana-d}
Deep, S., Koutris, P., Bidasaria, Y.: {QIRANA} demonstration: {R}eal time
  scalable query pricing.
\newblock PVLDB \textbf{10}(12), 1949--1952 (2017)

\bibitem{dwork2011firm}
Dwork, C.: A firm foundation for private data analysis.
\newblock Commun. ACM \textbf{54}(1), 86--95 (2011)

\bibitem{factual}
Factual: https://www.factual.com/

\bibitem{feng2019joint}
Feng, S., Niyato, D., Wang, P., Kim, D.I., Liang, Y.C.: Joint service pricing
  and cooperative relay communication for federated learning.
\newblock In: iThings/GreenCom/CPSCom/SmartData, pp. 815--820. IEEE (2019)

\bibitem{fernandez2020data}
Fernandez, R.C., Subramaniam, P., Franklin, M.J.: Data market platforms:
  Trading data assets to solve data problems.
\newblock PVLDB \textbf{13}(12), 1933--1947 (2020)

\bibitem{gal2011pricing}
Gal-Or, E.: Pricing practices of resellers in the airline industry: Posted
  price vs. name-your-own-price models.
\newblock Journal of economics \& management strategy \textbf{20}(1), 43--82
  (2011)

\bibitem{gbdex}
GBDEX: http://www.gbdex.com/website/

\bibitem{ge2005model}
Ge, W., Rothenberger, M., Chen, E.: A model for an electronic information
  marketplace.
\newblock Australasian Journal of Information Systems \textbf{13}(1) (2005)

\bibitem{ghorbani2019data}
Ghorbani, A., Zou, J.: Data shapley: Equitable valuation of data for machine
  learning.
\newblock In: ICML, pp. 2242--2251. PMLR (2019)

\bibitem{ghosh2011selling}
Ghosh, A., Roth, A.: Selling privacy at auction.
\newblock In: Proceedings of the 12th ACM conference on Electronic commerce,
  pp. 199--208 (2011)

\bibitem{glazer1993measuring}
Glazer, R.: Measuring the value of information: The information-intensive
  organization.
\newblock IBM Systems Journal \textbf{32}(1), 99--110 (1993)

\bibitem{gnip}
GNIP: http://support.gnip.com/apis/

\bibitem{google}
GoogleCloud: https://cloud.google.com/

\bibitem{heckman2015pricing}
Heckman, J.R., Boehmer, E.L., Peters, E.H., Davaloo, M., Kurup, N.G.: A pricing
  model for data markets.
\newblock iConference Proceedings  (2015)

\bibitem{here}
Here: https://www.here.com/en

\bibitem{hestness2017deep}
Hestness, J., Narang, S., Ardalani, N., Diamos, G., Jun, H., Kianinejad, H.,
  Patwary, M., Ali, M., Yang, Y., Zhou, Y.: Deep learning scaling is
  predictable, empirically.
\newblock arXiv preprint arXiv:1712.00409  (2017)

\bibitem{hu2020trading}
Hu, R., Gong, Y.: Trading data for learning: Incentive mechanism for on-device
  federated learning.
\newblock arXiv preprint arXiv:2009.05604  (2020)

\bibitem{hu2017optimal}
Hu, Z., Zhang, J.: Optimal posted-price mechanism in microtask crowdsourcing.
\newblock In: IJCAI, pp. 228--234. AAAI Press (2017)

\bibitem{ibm}
IBM: https://www.ibm.com/cloud/

\bibitem{infochimps}
InfoChimps: https://www.infochimps.com/

\bibitem{jia2019efficient}
Jia, R., Dao, D., Wang, B., Hubis, F.A., Gurel, N.M., Li, B., Zhang, C.,
  Spanos, C., Song, D.: Efficient task-specific data valuation for nearest
  neighbor algorithms.
\newblock PVLDB \textbf{12}(11), 1610--1623 (2019)

\bibitem{jiao2020toward}
Jiao, Y., Wang, P., Niyato, D., Lin, B., Kim, D.I.: Toward an automated auction
  framework for wireless federated learning services market.
\newblock IEEE Trans. Mob. Comput.  (2020, online)

\bibitem{jordan2019artificial}
Jordan, M.I.: Artificial intelligence—the revolution hasn’t happened yet.
\newblock Harvard Data Science Review \textbf{1}(1) (2019)

\bibitem{kandula2016quickr}
Kandula, S., Shanbhag, A., Vitorovic, A., Olma, M., Grandl, R., Chaudhuri, S.,
  Ding, B.: Quickr: Lazily approximating complex adhoc queries in bigdata
  clusters.
\newblock In: SIGMOD, pp. 631--646. ACM (2016)

\bibitem{koh2017understanding}
Koh, P.W., Liang, P.: Understanding black-box predictions via influence
  functions.
\newblock In: ICML, pp. 1885--1894. JMLR. org (2017)

\bibitem{koh2019accuracy}
Koh, P.W.W., Ang, K.S., Teo, H., Liang, P.S.: On the accuracy of influence
  functions for measuring group effects.
\newblock In: NeurIPS, pp. 5254--5264. Curran Associates, Inc. (2019)

\bibitem{koutris2012querymarket}
Koutris, P., Upadhyaya, P., Balazinska, M., Howe, B., Suciu, D.: Querymarket
  demonstration: Pricing for online data markets.
\newblock PVLDB \textbf{5}(12), 1962--1965 (2012)

\bibitem{koutris2013toward}
Koutris, P., Upadhyaya, P., Balazinska, M., Howe, B., Suciu, D.: Toward
  practical query pricing with querymarket.
\newblock In: SIGMOD, pp. 613--624. ACM (2013)

\bibitem{koutris2015query}
Koutris, P., Upadhyaya, P., Balazinska, M., Howe, B., Suciu, D.: Query-based
  data pricing.
\newblock Journal of the ACM \textbf{62}(5), 43 (2015)

\bibitem{kumar2018survey}
Kumar, D., Baranwal, G., Raza, Z., Vidyarthi, D.P.: A survey on spot pricing in
  cloud computing.
\newblock Journal of Network and Systems Management \textbf{26}(4), 809--856
  (2018)

\bibitem{li2014theory}
Li, C., Li, D.Y., Miklau, G., Suciu, D.: A theory of pricing private data.
\newblock ACM Trans. Database Syst. \textbf{39}(4), 1--28 (2014)

\bibitem{li2017theory}
Li, C., Li, D.Y., Miklau, G., Suciu, D.: A theory of pricing private data.
\newblock Commun. ACM \textbf{60}(12), 79--86 (2017)

\bibitem{li2012pricing}
Li, C., Miklau, G.: Pricing aggregate queries in a data marketplace.
\newblock In: WebDB, pp. 19--24. VLDB Endowment (2012)

\bibitem{li2018cost}
Li, Y., Sun, H., Dong, B., Wang, H.W.: Cost-efficient data acquisition on
  online data marketplaces for correlation analysis.
\newblock PVLDB \textbf{12}(4), 362--375 (2018)

\bibitem{li2013designing}
Li, Z., Li, B., Zhu, Y.: Designing truthful spectrum auctions for multi-hop
  secondary networks.
\newblock IEEE Trans. Mob. Comput. \textbf{14}(2), 316--327 (2013)

\bibitem{liang2018survey}
Liang, F., Yu, W., An, D., Yang, Q., Fu, X., Zhao, W.: A survey on big data
  market: Pricing, trading and protection.
\newblock IEEE Access \textbf{6}, 15,132--15,154 (2018)

\bibitem{lin2014arbitrage}
Lin, B.R., Kifer, D.: On arbitrage-free pricing for general data queries.
\newblock PVLDB \textbf{7}(9), 757--768 (2014)

\bibitem{liu2020dealer}
Liu, J.: Dealer: End-to-end data marketplace with model-based pricing.
\newblock arXiv preprint arXiv:2003.13103  (2020)

\bibitem{low2014graphlab}
Low, Y., Gonzalez, J.E., Kyrola, A., Bickson, D., Guestrin, C.E., Hellerstein,
  J.: Graphlab: A new framework for parallel machine learning.
\newblock arXiv preprint arXiv:1408.2041  (2014)

\bibitem{mcmahan2017communication}
McMahan, B., Moore, E., Ramage, D., Hampson, S., y~Arcas, B.A.:
  Communication-efficient learning of deep networks from decentralized data.
\newblock In: AISTATS, pp. 1273--1282. PMLR (2017)

\bibitem{miao2020towards}
Miao, X., Gao, Y., Chen, L., Peng, H., Yin, J., Li, Q.: Towards query pricing
  on incomplete data.
\newblock IEEE Trans. Knowl. Data Eng.  (2020, online)

\bibitem{azure}
MicrosoftAzure: http://datamarket.azure.com/

\bibitem{muschalle2012pricing}
Muschalle, A., Stahl, F., L{\"o}ser, A., Vossen, G.: Pricing approaches for
  data markets.
\newblock In: BIRTE, pp. 129--144. Springer (2012)

\bibitem{niu2019making}
Niu, C., Zheng, Z., Tang, S., Gao, X., Wu, F.: Making big money from small
  sensors: Trading time-series data under pufferfish privacy.
\newblock In: INFOCOM, pp. 568--576. IEEE (2019)

\bibitem{niu2020online}
Niu, C., Zheng, Z., Wu, F., Tang, S., Chen, G.: Online pricing with reserve
  price constraint for personal data markets.
\newblock In: ICDE, pp. 1978--1981. IEEE (2020)

\bibitem{niu2018unlocking}
Niu, C., Zheng, Z., Wu, F., Tang, S., Gao, X., Chen, G.: Unlocking the value of
  privacy: Trading aggregate statistics over private correlated data.
\newblock In: SIGKDD, pp. 2031--2040. ACM (2018)

\bibitem{niu2019erato}
Niu, C., Zheng, Z., Wu, F., Tang, S., Gao, X., Chen, G.: {ERATO}: Trading noisy
  aggregate statistics over private correlated data.
\newblock IEEE Trans. Knowl. Data Eng.  (2019, online)

\bibitem{north2010multiscale}
North, M.J., Macal, C.M., Aubin, J.S., Thimmapuram, P., Bragen, M., Hahn, J.,
  Karr, J., Brigham, N., Lacy, M.E., Hampton, D.: Multiscale agent-based
  consumer market modeling.
\newblock Complexity \textbf{15}(5), 37--47 (2010)

\bibitem{omnia}
Omnia: https://azuremarketplace.microsoft.com

\bibitem{pei2020survey}
Pei, J.: A survey on data pricing: from economics to data science.
\newblock arXiv preprint arXiv:2009.04462  (2020)

\bibitem{qilk}
Qilk: https://www.qlik.com/us/

\bibitem{qu2018posted}
Qu, Y., Tang, S., Dong, C., Li, P., Guo, S., Dai, H., Wu, F.: Posted pricing
  for chance constrained robust crowdsensing.
\newblock IEEE Trans. Mob. Comput. \textbf{19}(1), 188--199 (2018)

\bibitem{Reuters2016}
Reuters: Social media ad spending is expected to pass newspapers by 2020,
  http://fortune.com/2016/12/05/
  social-media-adspending-newspapers-zenith-2020/ (2016)

\bibitem{schomm2013marketplaces}
Schomm, F., Stahl, F., Vossen, G.: Marketplaces for data: An initial survey.
\newblock SIGMOD Record \textbf{42}(1), 15--26 (2013)

\bibitem{sen2013survey}
Sen, S., Joe-Wong, C., Ha, S., Chiang, M.: A survey of smart data pricing: Past
  proposals, current plans, and future trends.
\newblock ACM Computing Surveys (CSUR) \textbf{46}(2), 15 (2013)

\bibitem{shapiro1998versioning}
Shapiro, C., Varian, H.R.: Versioning: The smart way to sell information.
\newblock Harvard Business Review \textbf{107}(6), 107 (1998)

\bibitem{shen2019meal}
Shen, Z., He, Z., Xue, X.: Meal: Multi-model ensemble via adversarial learning.
\newblock In: AAAI, vol.~33, pp. 4886--4893. Association for the Advancement of
  Artificial Intelligence (2019)

\bibitem{syrgkanis2015pricing}
Syrgkanis, V., Gehrke, J.: Pricing queries approximately optimally.
\newblock arXiv preprint arXiv:1508.05347  (2015)

\bibitem{ruiming2014quality}
Tang, R.: On the quality and price of data.
\newblock Ph.D. thesis, National University of Singapore (2014)

\bibitem{tang2014get}
Tang, R., Amarilli, A., Senellart, P., Bressan, S.: Get a sample for a discount
  sampling-based {XML} data pricing.
\newblock In: DEXA, pp. 20--34. Springer (2014)

\bibitem{tang2013you}
Tang, R., Shao, D., Bressan, S., Valduriez, P.: What you pay for is what you
  get.
\newblock In: DEXA, pp. 395--409. Springer (2013)

\bibitem{tang2013price}
Tang, R., Wu, H., Bao, Z., Bressan, S., Valduriez, P.: The price is right -
  models and algorithms for pricing data.
\newblock In: DEXA, pp. 380--394. Springer (2013)

\bibitem{twitter}
Twitter: https://twitter.com/

\bibitem{upadhyaya2015managing}
Upadhyaya, P.: Managing premium data.
\newblock Ph.D. thesis (2015)

\bibitem{upadhyaya2012price}
Upadhyaya, P., Balazinska, M., Suciu, D.: How to price shared optimizations in
  the cloud.
\newblock PVLDB \textbf{5}(6), 562--573 (2012)

\bibitem{upadhyaya2016price}
Upadhyaya, P., Balazinska, M., Suciu, D.: Price-optimal querying with data
  {API}s.
\newblock PVLDB \textbf{9}(14), 1695--1706 (2016)

\bibitem{vagata2014scaling}
Vagata, P., Wilfong, K.: Scaling the facebook data warehouse to 300 {PB}.
\newblock Facebook Code, Facebook \textbf{10} (2014)

\bibitem{wang2012cloud}
Wang, Q., Ren, K., Meng, X.: When cloud meets ebay: {T}owards effective pricing
  for cloud computing.
\newblock In: INFOCOM, pp. 936--944. IEEE (2012)

\bibitem{wang2019novel}
Wang, X., Wei, X., Gao, S., Liu, Y., Li, Z.: A novel auction-based query
  pricing schema.
\newblock International Journal of Parallel Programming \textbf{47}, 1--22
  (2019)

\bibitem{wang2018pricing}
Wang, X., Wei, X., Liu, Y., Gao, S.: On pricing approximate queries.
\newblock Inf. Sci. \textbf{453}, 198--215 (2018)

\bibitem{xignite}
Xignite: https://www.xignite.com/

\bibitem{xiong2019smart}
Xiong, W., Xiong, L.: Smart contract based data trading mode using blockchain
  and machine learning.
\newblock IEEE Access \textbf{7}, 102,331--102,344 (2019)

\bibitem{yang2020game}
Yang, J., Fan, J., Wei, Z., Li, G., Liu, T., Du, X.: A game-based framework for
  crowdsourced data labeling.
\newblock VLDB J.  (2020, online)

\bibitem{yoon2019data}
Yoon, J., Arik, S.O., Pfister, T.: Data valuation using reinforcement learning.
\newblock In: ICML. PMLR (2020, online)

\bibitem{youedata}
Youedata: https://www.youedata.com/

\bibitem{yu2017data}
Yu, H., Zhang, M.: Data pricing strategy based on data quality.
\newblock Comput. Ind. Eng. \textbf{112}, 1--10 (2017)

\bibitem{zhang2020multi}
Zhang, C., Jiang, M., Zhang, X., Ye, Y., Chawla, N.V.: Multi-modal network
  representation learning.
\newblock In: SIGKDD, pp. 3557--3558. ACM (2020)

\bibitem{zhang2016materialization}
Zhang, C., Kumar, A., R{\'e}, C.: Materialization optimizations for feature
  selection workloads.
\newblock ACM Trans. Database Syst. \textbf{41}(1), 2 (2016)

\bibitem{zhang2014dynamic}
Zhang, L., Li, Z., Wu, C.: Dynamic resource provisioning in cloud computing: A
  randomized auction approach.
\newblock In: INFOCOM, pp. 433--441. IEEE (2014)

\bibitem{Zhang2020survey}
Zhang, M., Beltrán, F.: A survey of data pricing methods,
  https://ssrn.com/abstract=3609120 (2020)

\bibitem{zheng2017online}
Zheng, Z., Peng, Y., Wu, F., Tang, S., Chen, G.: An online pricing mechanism
  for mobile crowdsensing data markets.
\newblock In: MobiHoc, pp. 1--10. ACM (2017)

\bibitem{zheng2017trading}
Zheng, Z., Peng, Y., Wu, F., Tang, S., Chen, G.: Trading data in the crowd:
  Profit-driven data acquisition for mobile crowdsensing.
\newblock IEEE Journal on Selected Areas in Communications \textbf{35}(2),
  486--501 (2017)

\bibitem{zheng2019arete}
Zheng, Z., Peng, Y., Wu, F., Tang, S., Chen, G.: {Arete}: On designing joint
  online pricing and reward sharing mechanisms for mobile data markets.
\newblock IEEE Trans. Mob. Comput. \textbf{19}(4), 769--787 (2019)

\bibitem{Zhu2020snpricing}
Zhu, Y., Tang, J., Tang, X.: Pricing influential nodes in online social
  networks.
\newblock PVLDB \textbf{13}(10), 1614--1627 (2020)

\end{thebibliography}
}

\end{document}